\documentclass[reprint,aps,prl,superscriptaddress,notitlepage]{revtex4-2}
\usepackage{amssymb,amsmath}
\usepackage{graphicx}
\usepackage{color}
\usepackage[english]{babel}
\usepackage[caption=false]{subfig}
\usepackage[colorlinks=true,allcolors=blue]{hyperref}
\usepackage{braket}
\usepackage{xr}
\begin{document} 

\def\simlt{\mathrel{\lower .3ex \rlap{$\sim$}\raise .5ex \hbox{$<$}}}

\title{\textbf{\fontfamily{phv}\selectfont How valley-orbit states in silicon quantum dots probe quantum well interfaces}}
% Tunable valley-orbit states in Si/SiGe quantum dots probe heterostructure interface
\author{J. P. Dodson}
\affiliation{Department of Physics, University of Wisconsin-Madison, Madison, WI 53706, USA}
\author{H. Ekmel Ercan}
\affiliation{Department of Physics, University of Wisconsin-Madison, Madison, WI 53706, USA}
\author{J. Corrigan}
\affiliation{Department of Physics, University of Wisconsin-Madison, Madison, WI 53706, USA}
\author{Merritt P. Losert}
\affiliation{Department of Physics, University of Wisconsin-Madison, Madison, WI 53706, USA}
\author{Nathan Holman}
\affiliation{Department of Physics, University of Wisconsin-Madison, Madison, WI 53706, USA}
\author{Thomas McJunkin}
\affiliation{Department of Physics, University of Wisconsin-Madison, Madison, WI 53706, USA}
\author{L. F. Edge}
\affiliation{HRL Laboratories, LLC, 3011 Malibu Canyon Road, Malibu, CA 90265, USA}
\author{Mark Friesen}
\affiliation{Department of Physics, University of Wisconsin-Madison, Madison, WI 53706, USA}
\author{S. N. Coppersmith}
\affiliation{Department of Physics, University of Wisconsin-Madison, Madison, WI 53706, USA}
\affiliation{University of New South Wales, Sydney, Australia}|
\author{M. A. Eriksson}
\affiliation{Department of Physics, University of Wisconsin-Madison, Madison, WI 53706, USA}

\begin{abstract}
The energies of valley-orbit states in silicon quantum dots are determined by an as yet poorly understood interplay between interface roughness, orbital confinement, and  electron interactions.  Here, we report measurements of one- and two-electron valley-orbit state energies as the dot potential is modified by changing gate voltages, and we calculate these same energies using full configuration interaction calculations. The results enable an understanding of the interplay between the physical contributions and enable a new probe of the quantum well interface.
\end{abstract}

\maketitle
The ability to make uniform and tunable qubits is crucial for large-scale applications.  Modern computers use one control electrode per field effect transistor with excellent uniformity, and proposed architectures for quantum chips also rely on a small number of control lines per qubit, in order to minimize the density of control wires~\cite{Vandersypen:2017p1,Veldhorst:2017p1}. Progress has recently been made enhancing the homogeneity of the electrical environment by using quantum dot designs that eliminate modulation doping and instead make use of metal surface electrodes to both accumulate and deplete electrons~\cite{Zajac:2016p054013}.  The resulting structures enable good control over electron occupation, gate voltages, and tunnel couplings between quantum dots with a small number of gate electrodes per quantum dot~\cite{Zajac:2018p439, Mi:2018p599, Neyens:2019p08216, Dodson:2020p505001, Mills:2019p1, Andrews:2019p05004, Holman:2020p083502, Takeda:2021p41565}.

Uniformity remains a challenge with regards to conduction band valley energies in silicon~\cite{Ando:1982p437,Boykin:2004p115}, and important physical questions need to be addressed. Atomistic disorder is known to play a particularly important (and typically uncontrolled) role in determining the energies of electrons at the bottom of the valleys \cite{Friesen:2006p202106, Kharche:2007p092109, Culcer:2010p205315, Gamble:2013p035310, Boross:2016p035438, Hosseinkhani:2020p043180}, resulting in a wide range of observed valley splittings in Si/SiGe quantum dots (20--270 $\mu$eV, \cite{Shaji:2008p540, Simmons:2010p245312, Shi:2011p233108, Borselli:2011p123118, Kim:2014p70, Schoenfield:2017p64, Mi:2018p76803, Jones:2019p014026, Hollmann:2020p34068, Chen:2020p44033}).  Critically, the interplay between the factors that determine the valley splitting in quantum dots --- the atomic details of the interface (which vary with lateral position), the degree of lateral confinement, and electron-electron interactions within a quantum dot --- are not yet fully understood.

\begin{figure}[b]
\includegraphics[width=0.47\textwidth]{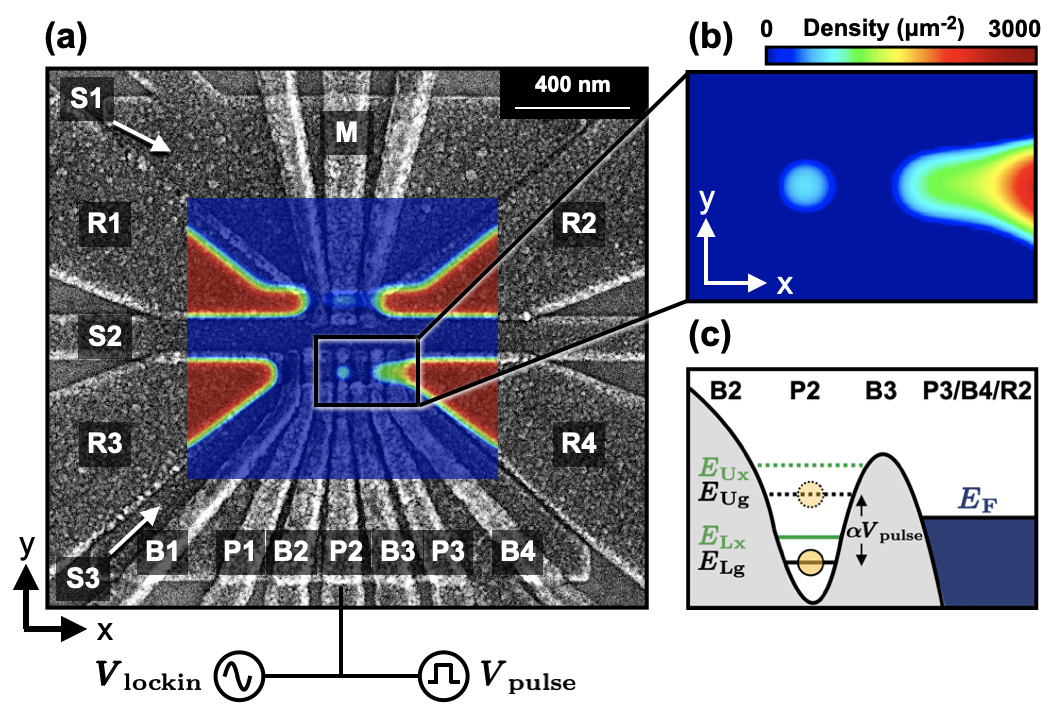}
\caption{\label{fig:fig1} Device layout and experimental setup. (a) A scanning electron microscope (SEM) image of a device lithographically identical to the one measured shows the gate electrode layout in the active region. A COMSOL Multiphysics Thomas-Fermi simulation of the electron density is overlaid on the SEM image. The sensor dot under gate M measures the average charge occupation $\braket{n}$ of the P2 quantum dot via lock-in amplifier detection at frequency $f_{\text{lockin}}$. (b) The COMSOL simulation shows the charge density of the P2 dot, where the tunnel barrier to the left(right) reservoir is opaque(transparent). (c) Valley-orbit state splittings are measured using pulsed-gate spectroscopy. A 50\% duty cycle square pulse with amplitude $V_{\text{pulse}}$ at frequency $f_{\text{pulse}}$ is applied to gate P2, rapidly pulsing the chemical potential of the P2 dot between $E_{\text{Lg}}$ and $E_{\text{Ug}}$. The change in chemical potential induces detectable shifts in the tunnel rate into and out of the P2 quantum dot, allowing for measurement of excited state energies.}
\end{figure} 
This Letter reports quantitative characterization of the relationship between low-lying one- and two-electron valley-orbit states and the quantum dot confinement strength, shape, and position. The pulsed-gate spectroscopy and magnetospectroscopy measurements reveal valley splittings in the range 36--87 $\mu$eV, two-electron singlet-triplet splittings between 22--59 $\mu$eV, and orbital splittings that can be tuned from 1.69--2.26 meV. Simulations combining full configuration interaction (FCI) \cite{Szabo:1996p91861} with empirical tight-binding (TB) theory \cite{Boykin:2004p165325} are shown to be in good agreement with the experimental results, and together these methods enable an understanding of the interplay between effects arising from quantum well interface roughness, orbital confinement strength, and  electron-electron (e-e) interactions. This combination of experiment and theory not only explains the origin of the energy spectrum but also provides a new method for probing the quantum well interface.

Spectroscopy of one- and two-electron valley-orbit states is performed in a device fabricated using a three-layer overlapping aluminum gate architecture \cite{Zajac:2016p054013}, as shown in Fig.~\ref{fig:fig1}(a). A detailed fabrication process can be found in Ref.\ \cite{Dodson:2020p505001}. The integrated sensor dot under gate M measures the electron occupation of the central quantum dot under gate P2. The triple-dot on the bottom side is tuned into a regime where B1 and P1 form a large tunnel barrier on the left side of P2, suppressing the tunnel rate into reservoir R3. Gates P3 and B4 extend the reservoir R4 into the quantum dot channel, as shown by the electron density heat map in Fig.~\ref{fig:fig1}(b), allowing for the tunnel barrier beneath gate B3 to tune the tunnel coupling between the P2 dot and right reservoir under R4. Screening gates S1, S2 and S3 control the y-confinement of the P2 dot. The electron temperature is measured to be $T_e = 100$ mK.

One- and two-electron valley-orbit splittings are measured  by pulsed-gate spectroscopy using the experimental setup shown in Fig.~\ref{fig:fig1}(c). A 50\% duty cycle square voltage pulse with amplitude $V_{\text{pulse}}$ and frequency $f_{\text{pulse}}$ is applied to gate P2, pulsing the ground state of the quantum dot between two levels: $E_{\text{Lg}}$ and $E_{\text{Ug}} = E_\text{Lg} + \alpha V_{\text{pulse}}$, where $E_{\text{Lg}}$($E_{\text{Ug}}$) denotes the ground state of the dot in the loading(unloading) position, and $\alpha$ is the lever arm for gate P2. A measurable change in the average electron occupation of the dot $\braket{n}$ occurs when an excited state provides an additional channel for the electron to enter the dot, yielding a measurement of the energy of this excited state \cite{Elzerman:2004p731, Yang:2012p15319, Gamble:2016p253101}.

\begin{figure}[t]
\includegraphics[width=0.48\textwidth]{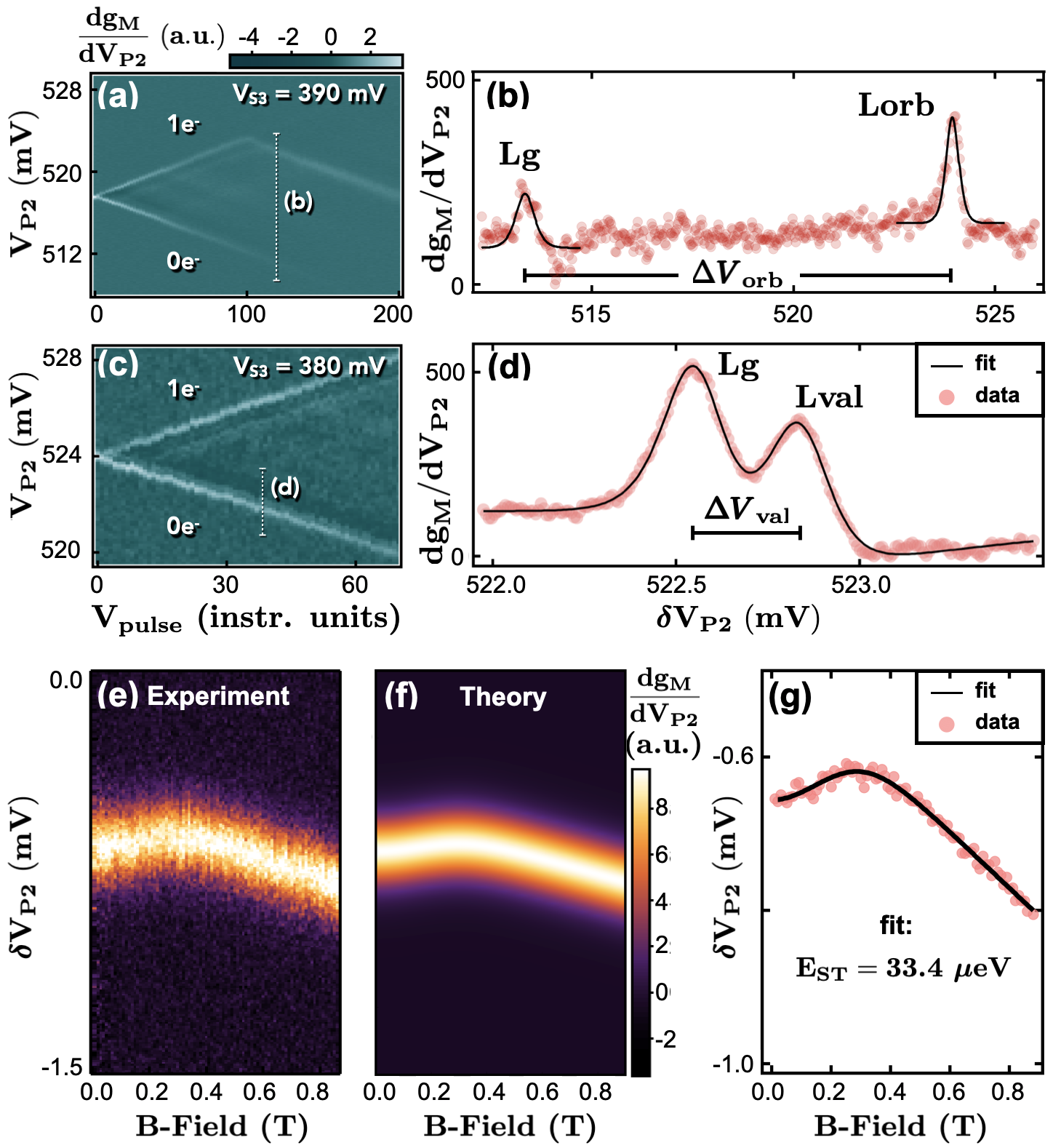}
\caption{\label{fig:fig2} Experimental methods. (a, b) Measurement of orbital splitting using pulsed-gate spectroscopy. The pulse amplitude is increased until the lowest-lying orbital state is visible, shown in (a). A line cut (dashed line) is taken when the pulse amplitude exceeds the orbital splitting, allowing detection of the ground (Lg) and excited orbital (Lorb) states, shown as the peaks in (b). The orbital splitting is given by $\alpha \Delta V_{\text{orb}}$. (c, d) Using the same method, the valley ($E_{\text{val}}$) and singlet-triplet splittings ($E_{\text{ST}}$) are measured. The red data points clearly show two distinct peaks in (d). The data is fit using Eqn.~\ref{eq:navg_pulsed_gate_final_2}, shown as the black line, giving $E_{\text{val}} = \alpha \Delta V_{\text{val}} = 53.2$ $\mu$eV. \mbox{(e-g)} Experimental magnetospectroscopy data (e) is reproduced in (f) by treating the dot-reservoir system as a grand canonical ensemble. The experimental peak locations in voltage space are extracted from (e) and plotted in (g) as red circles, which are then fit to Eqn~\ref{eq:fitVg_2}, yielding $E_{\text{ST}}=33.4$ $\mu$eV.}
\end{figure}
Fig.~\ref{fig:fig2}(a, b) show pulsed-gate spectroscopy measurements of the one-electron $E_{\text{orb}}$ at $V_{\text{S3}}=390$ mV. Differential conductance measurements, reported as $dg_{\text{M}}/dV_{\text{P2}}$ are used to determine the location of excited states at particular gate voltages. The excited orbital state is separated well enough from the ground state such that each peak position is found by fitting to the derivative of a Fermi function \cite{Field:1993p1477, Kouwenhoven:1997p1384}, where $E_{\text{orb}} = \alpha \Delta V_{\text{orb}}$. As shown in Fig.~\ref{fig:fig2}(c, d), a similar procedure is used to measure the valley splitting at $V_{\text{S3}}=380$ mV, which is easily differentiated from the orbital splitting since it is much lower-lying in energy. In this case, there is overlap of the ground and lowest excited state signals that arises from thermal broadening.  To extract the peak locations, we make use of an expression for $\braket{n}$,
\begin{equation}
\label{eq:navg_pulsed_gate_final_2}
\braket{n} = \sum \limits_{i = g,x} \Gamma_i \dfrac{e^{(E_i-E_{F})/E_{0i}}}{e^{(E_i-E_F)/k_B T_e} + 1},
\end{equation}
%\braket{n_{b}} = \dfrac{T_b}{4}\sum \limits_{i = g, x} f_{bi}\Gamma_{bi}
\noindent where $E_i = \alpha V_i$ is the position of each peak in energy, and $E_{0i}$ and $\Gamma_i$ are fitting parameters for each peak.
As shown by the solid line in Fig.~2(d), the experimental data are fit by the derivative of Eqn.~\ref{eq:navg_pulsed_gate_final_2} with respect to the gate voltage ($d\braket{n}/dV_{\text{P2}}$), enabling extraction of $E_{\text{val}} = \alpha \Delta V_\text{val}$. 

The two-electron singlet-triplet splitting ($E_{\text{ST}}$) is measured using both pulsed-gate spectroscopy (using a similar procedure as above) and magnetospectroscopy, as shown in Fig.~\ref{fig:fig2}(e-g). The latter is performed by adiabatically sweeping across the $1\rightarrow 2$ charge transition line and measuring the resulting position of the charge transition peak. Measurements of $E_{\text{ST}}$ using each of these methods are shown to be in agreement in Section \ref{magpulse} of the Supplementary Material. This method is described in further detail in Refs.~\cite{Lim:2009p242102, Lim:2011p35704, Borselli:2011p123118, Shi:2011p233108}. A full theoretical model for magnetospectroscopy is developed in the Supplemental Material which allows for the experimental data shown in Fig.~\ref{fig:fig2}(e) to be closely reproduced by the model in Fig.~\ref{fig:fig2}(f). This model enables fitting the peak position of the data in Fig.~\ref{fig:fig2}(e), using
\begin{align}
 V_{P2}(B) = \dfrac{1}{\alpha\beta_e} \ln \left(\dfrac{e^{\frac{1}{2} \kappa B + \beta_e E_{\text{ST}}}\left(e^{\kappa B}+1\right)}{e^{\kappa B}+e^{2 \kappa B}+e^{\kappa B + \beta_e E_{\text{ST}}}+1}\right), \label{eq:fitVg_2}
\end{align}
where $V_{P2}$ is the gate voltage, $\kappa=g \mu_B \beta_e$ where $\beta_e=1/k_BT_e$, $g$ is the electron g-factor, $\mu_B$ is the Bohr magneton, and $B$ is the magnetic field. The peak positions from Fig.~\ref{fig:fig2}(e) are extracted and plotted as red circles in Fig.~\ref{fig:fig2}(g), and the fit (solid line)  to Eqn.~\ref{eq:fitVg_2} yields for this example $E_{\text{ST}} = 33.4$ $\mu$eV. 

We now show that by combining these two different spectroscopic techniques, and by combining two different theoretical techniques, we can extract a quantitative measure of the suppression of the singlet-triplet splitting from the valley splitting. Further, below we show that these same techniques allow us to extract information about the atomic details of the interface by comparing measurement of $E_{\text{val}}$, $E_{\text{ST}}$ and $E_{\text{orb}}$ vs. electrostatic confinement in the x-y plane. The confinement is varied by changing the S3 gate voltage ($V_{\text{S3}}$) between 260--420 mV while compensating with neighboring barrier/plunger gates to maintain a constant electron occupation and tunnel rate into the dot.

\begin{figure}
\includegraphics[width=0.48\textwidth]{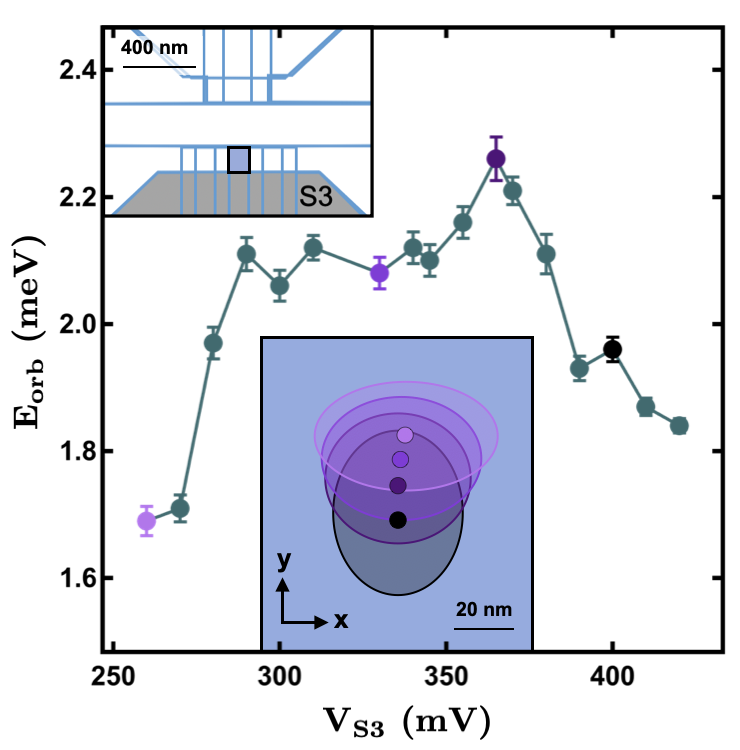}
\caption{\label{fig:fig3} Quantum dot orbital shape and position. The top left inset shows a device schematic where the P2 quantum dot is located within the filled blue region and the S3 gate is shown as the filled gray region. The minimum orbital splitting is plotted as a function of $V_{\text{S3}}$. At low $V_{\text{S3}}$, the dot has strong y-confinement and weak x-confinement, leading to a drop in the minimum orbital splitting. As $V_{\text{S3}}$ is increased, the dot becomes isotropic, increasing the minimum orbital splitting. At high $V_{\text{S3}}$, the dot has weak y-confinement and strong x-confinement, reducing the minimum orbital splitting again. Thomas-Fermi electrostatic simulations are used to calculate the quantum dot shape and position within the blue region shown in the device schematic. Four points are simulated, shown as the shaded purple circles in the experimental data. The change in shape of the electron density is found to be in agreement with the experimental data. Additionally, the position of the quantum dot for each simulation is illustrated as the small solid circle in the inset, showing that the quantum dot slides down and to the left 23.8 nm.}
\end{figure}

Figure~\ref{fig:fig3} shows the effects of changing $V_{\text{S3}}$ on quantum dot orbital energy, shape, and position. The device schematic pictured in the top left inset shows the approximate location of the P2 dot as the shaded blue region. Changes to the voltage $V_{\text{S3}}$ applied to the screening gate, shown as the shaded gray region in the inset, modify the P2 dot confinement, orbital shape, and position. The minimum orbital splitting, plotted as solid circles in Figure~\ref{fig:fig3}, is found to be non-monotonic with $V_{\text{S3}}$, because of the strong effect S3 has on the electrostatic confinement of the dot along the y-axis. At high $V_{\text{S3}}$, the y-confinement becomes weak, and the minimum orbital splitting falls off rapidly due to elongation of the dot along the y-axis. Towards the center, the dot becomes isotropic, increasing the minimum orbital splitting up to 2.26 meV at $V_{\text{S3}}=370$ mV. At low $V_{\text{S3}}$, suppression of the minimum orbital splitting occurs due to the compensating barrier and plunger gate voltages needed to stay in the one-electron regime, elongating the dot along the x-axis. Since low $V_{\text{S3}}$ creates a tight confinement potential along the y-axis, the x-orbital is the most weakly confined orbital below $V_{\text{S3}}=370$ mV.

This behavior is well explained by semiclassical (Thomas-Fermi) electrostatic simulations using COMSOL Multiphysics, shown in the bottom inset of Fig.~\ref{fig:fig3}. The position and orbital shape are simulated at the four points indicated by the shaded purple circles from the experimental data. The outline of each oval represents the shape that encloses 50\% of the electron wave function, and the small circles show the center of the electron density for each simulation. The Thomas-Fermi (TF) simulations qualitatively match the experimental findings, where the x-orbital is most weakly confined at low $V_{\text{S3}}$ and the y-orbital is most weakly confined at high $V_{\text{S3}}$. In addition to the changes in shape, significant change in the position of the dot is observed in the COMSOL simulations. Over the range shown, the quantum dot position slides down and to the left as $V_{\text{S3}}$ increases for a total change in position of 23.8 nm. We note that COMSOL simulations were not used to determine the dot shape and position above 400 mV, because this regime becomes close to the accumulation threshold for the S3 gate, where uncertainties become large.

\begin{figure}
\includegraphics[width=0.46\textwidth]{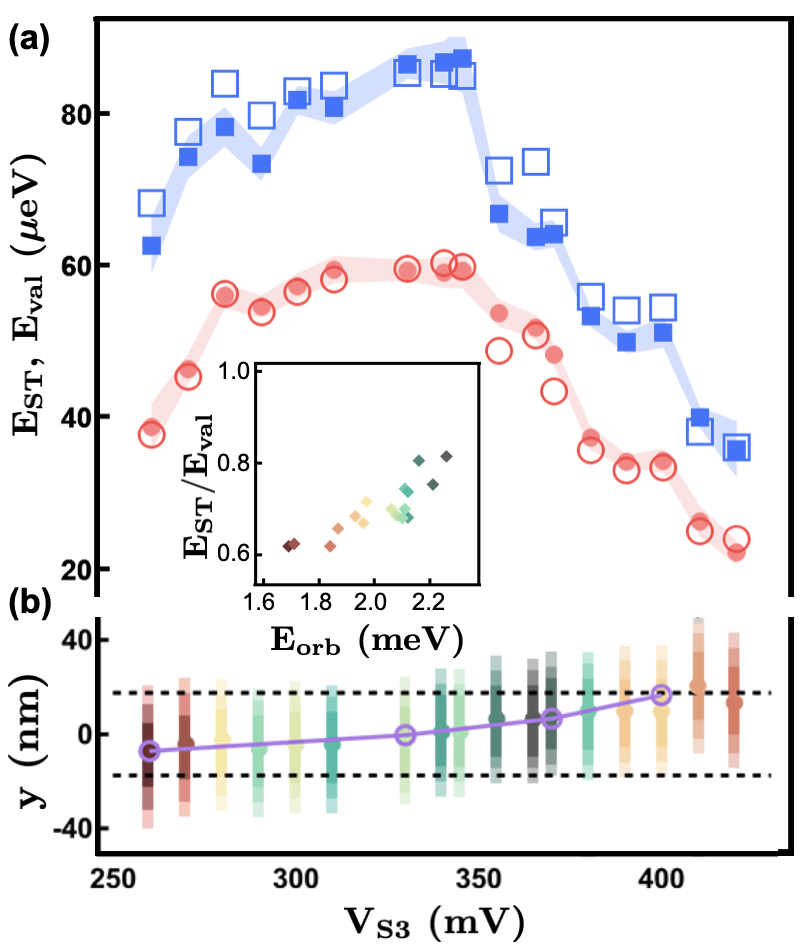}
\caption{\label{fig:fig4} (a) The measured valley splitting ($E_{\text{val}}$) and singlet-triplet splitting ($E_{\text{ST}}$) are plotted as a function of $V_{\text{S3}}$ as filled blue squares and red circles, respectively, where the measurement uncertainty is indicated by the width of the shaded regions. FCI simulations for $E_{\text{val}}$ and $E_{\text{ST}}$, shown as open blue squares and red circles, respectively, quantitatively reproduce the experimental data once the effects of disorder in the form of atomic steps at the quantum well interface are included. (b) The non-monotonic behavior observed in both $E_{\text{ST}}$ and $E_{\text{val}}$ as a function $V_{\text{S3}}$ is well explained by the quantum dot position changing with respect to distinct atomic steps at the quantum well interface. The wave function and position of the P2 dot singlet state is plotted with respect to atomic steps (dashed lines) at the interface, shown as the colored gradients and points. A best fit to the data produces a smooth change in position of the dot across steps spaced approximately 35 nm apart. The change in position from the Thomas-Fermi simulation, plotted as the purple line and open circles, agrees well with the FCI simulations. The inset plots the ratio $E_{\text{ST}}/E_{\text{val}}$ as a function of the orbital splitting ($E_{\text{orb}}$), where the colored points map the magnitude of the orbital splitting to the position of the electron wave function with respect to atomic steps in (b). As seen in (b), the suppression of the ratio $E_{\text{ST}}/E_{\text{val}}$ below unity is a consequence of strong electron-electron interactions dominated by the magnitude of the orbital splitting.}
\end{figure}

Atomic steps in the quantum well interface play an important role in determining $E_{\text{val}}$ and $E_{\text{ST}}$. Fig.~\ref{fig:fig4}(a) plots $E_{\text{val}}$ (filled blue squares) and $E_{\text{ST}}$ (filled red circles) as the orbital shape and dot position are varied with $V_{\text{S3}}$, where the shaded region represents the measurement uncertainty. The measured $E_{\text{val}}$ and $E_{\text{ST}}$ change substantially across the electrostatic configurations, and their large \textit{in situ} tunability arises from motion of the P2 quantum dot with respect to atomic steps at the quantum well interface. As the dot approaches an atomic step, more of the wave function overlaps the step, thereby increasing the valley-orbit coupling which suppresses $E_{\text{val}}$ and $E_{\text{ST}}$ \cite{Friesen:2010p115324}. To determine the position of atomic steps relative to the quantum dot, we make use of a combination of the experimental measurements, the COMSOL simulations just described, and FCI calculations, the latter of which incorporate valley-orbit coupling that arises from interface roughness. The FCI calculations use a single fitting parameter in order to match the measured values of $E_{\text{val}}$ and $E_{\text{ST}}$ namely, the position of the dot relative to atomic steps at the quantum well interface.

Figure~\ref{fig:fig4}(a) shows the results of FCI calculations for $E_{\text{val}}$ and $E_{\text{ST}}$ as open blue squares and open red circles, respectively. Close agreement with the experimental data is found when the atomic steps are separated by 35 nm. The positions used in the FCI calculations to produce the best fit to the data at each point are plotted in Fig.~\ref{fig:fig4}(b) as solid colored points. Each point has a colored gradient overlay that represents the spatial extent of the singlet (ground) state with respect to the atomic steps, which are shown as black dashed lines. The gradient steps from darkest to lightest, corresponding to wave function probability thresholds of 75\%, 50\%, and 25\% of its maximum value.

A linear fit to the dot positions calculated from the FCI simulations reveal a total change in position of the dot of 24.7 nm with respect to the axis perpendicular to the atomic steps in the interface, which is consistent with the change in position extracted from the COMSOL simulations: 23.8 nm. This correspondence is shown in Fig.~\ref{fig:fig4}(b), which shows the dot position from each of the four COMSOL simulations (purple open circles) overlaid on the calculated FCI wavefunction distributions (color gradients). The close agreement provides a measure of validation for an approach that allows an \textit{in situ} method of probing the quantum well interface through a combination of spectroscopic measurements of the quantum dot and theoretical simulations. This is an important development since currently the only method of probing atomic details of buried interfaces are destructive metrology techniques such as transmission electron microscopy (TEM), nanobeam X-ray, or atom probe tomography (APT) \cite{McJunkin:2021p85406, Tilka:2016p55043, Osti:2017p74844}. We note that in future work, more detailed information could be determined by moving the dot along two axes. In this experiment, the direction of the atomic steps with respect to the movement vector of the quantum dot cannot be determined.

The inset of Fig.~\ref{fig:fig4}(a) reports the ratio $E_{\text{ST}}/E_{\text{val}}$ as a function of $E_{\text{orb}}$, showing that this ratio is significantly below unity for the entire range of parameters measured.  Electron-electron interactions suppress $E_{\text{ST}}$ below the non-interacting energy $E_{\text{val}}$ \cite{Pecker:2013p2692, Ercan:2021p35302} revealing that there are strong e-e interactions across all orbital splittings studied here (1.69--2.26 meV). To look for any correlations between dot position relative to the steps and the ratio $E_{\text{ST}}/E_{\text{val}}$, the dot positions in Fig.~\ref{fig:fig4}(b) are color coded by their corresponding orbital energies, with the mapping shown in the inset plot. The colors of the data points map the magnitude of the orbital splitting to the distance of the dot from an atomic step. The random distribution of colors in Fig.~\ref{fig:fig4}(b) demonstrates that there is very little correlation between the the distance of the dot from an atomic step and the orbital splitting. Thus, the orbital splitting is more important than dot position in determining the suppression of $E_\text{ST}$ below its non-interacting limit of $E_{\text{val}}$.

In summary, we measure large \textit{in situ} tunability of valley, singlet-triplet, and orbital splittings, allowing for determination of the quantitative relationship between these three important energy scales. FCI simulations of the measured valley-orbit states were found to be in good agreement with the data, pointing to a combination of two primary physical parameters driving the relationship between these valley-orbit states: electrostatic confinement strength and quantum dot position relative to steps in the quantum well interface. This improved understanding enables a new \textit{in situ} method of probing the quantum well interface through a combination of spectroscopic data and theoretical simulations.

The derivation of Eqns.~1 and 2, details of the COMSOL simulations, and details of the FCI calculations are provided in the Supplemental Material \cite{Penthorn:2020p08680, MacLean:2007p1499, Dicarlo:2004p1440, Hanson:2007p1217, vanderWiel:2003p1, Elzerman:2003p728, Kittel:1980p10882, Stopa:1996p13767, Wuetz:2020p02305}.

\begin{acknowledgments}
Research was sponsored in part by the Army Research Office (ARO) under Grant Numbers W911NF-17-1-0274 and by the Vannevar Bush Faculty Fellowship program under ONR grant number N00014-15-1-0029. We acknowledge the use of facilities supported by NSF through the UW-Madison MRSEC (DMR-1720415) and the NSF MRI program (DMR-1625348). The views and conclusions contained in this document are those of the authors and should not be interpreted as representing the official policies, either expressed or implied, of the Army Research Office (ARO), or the U.S. Government. The U.S. Government is authorized to reproduce and distribute reprints for Government purposes notwithstanding any copyright notation herein.
\end{acknowledgments}

\bibliography{main.bib}

\begin{thebibliography}{53}%
\makeatletter
\providecommand \@ifxundefined [1]{%
 \@ifx{#1\undefined}
}%
\providecommand \@ifnum [1]{%
 \ifnum #1\expandafter \@firstoftwo
 \else \expandafter \@secondoftwo
 \fi
}%
\providecommand \@ifx [1]{%
 \ifx #1\expandafter \@firstoftwo
 \else \expandafter \@secondoftwo
 \fi
}%
\providecommand \natexlab [1]{#1}%
\providecommand \enquote  [1]{``#1''}%
\providecommand \bibnamefont  [1]{#1}%
\providecommand \bibfnamefont [1]{#1}%
\providecommand \citenamefont [1]{#1}%
\providecommand \href@noop [0]{\@secondoftwo}%
\providecommand \href [0]{\begingroup \@sanitize@url \@href}%
\providecommand \@href[1]{\@@startlink{#1}\@@href}%
\providecommand \@@href[1]{\endgroup#1\@@endlink}%
\providecommand \@sanitize@url [0]{\catcode `\\12\catcode `\$12\catcode
  `\&12\catcode `\#12\catcode `\^12\catcode `\_12\catcode `\%12\relax}%
\providecommand \@@startlink[1]{}%
\providecommand \@@endlink[0]{}%
\providecommand \url  [0]{\begingroup\@sanitize@url \@url }%
\providecommand \@url [1]{\endgroup\@href {#1}{\urlprefix }}%
\providecommand \urlprefix  [0]{URL }%
\providecommand \Eprint [0]{\href }%
\providecommand \doibase [0]{https://doi.org/}%
\providecommand \selectlanguage [0]{\@gobble}%
\providecommand \bibinfo  [0]{\@secondoftwo}%
\providecommand \bibfield  [0]{\@secondoftwo}%
\providecommand \translation [1]{[#1]}%
\providecommand \BibitemOpen [0]{}%
\providecommand \bibitemStop [0]{}%
\providecommand \bibitemNoStop [0]{.\EOS\space}%
\providecommand \EOS [0]{\spacefactor3000\relax}%
\providecommand \BibitemShut  [1]{\csname bibitem#1\endcsname}%
\let\auto@bib@innerbib\@empty
%</preamble>
\bibitem [{\citenamefont {Vandersypen}\ \emph {et~al.}(2017)\citenamefont
  {Vandersypen}, \citenamefont {Bluhm}, \citenamefont {Clarke}, \citenamefont
  {Dzurak}, \citenamefont {Ishihara}, \citenamefont {Morello}, \citenamefont
  {Reilly}, \citenamefont {Schreiber},\ and\ \citenamefont
  {Veldhorst}}]{Vandersypen:2017p1}%
  \BibitemOpen
  \bibfield  {author} {\bibinfo {author} {\bibfnamefont {L.}~\bibnamefont
  {Vandersypen}}, \bibinfo {author} {\bibfnamefont {H.}~\bibnamefont {Bluhm}},
  \bibinfo {author} {\bibfnamefont {J.}~\bibnamefont {Clarke}}, \bibinfo
  {author} {\bibfnamefont {A.}~\bibnamefont {Dzurak}}, \bibinfo {author}
  {\bibfnamefont {R.}~\bibnamefont {Ishihara}}, \bibinfo {author}
  {\bibfnamefont {A.}~\bibnamefont {Morello}}, \bibinfo {author} {\bibfnamefont
  {D.}~\bibnamefont {Reilly}}, \bibinfo {author} {\bibfnamefont
  {L.}~\bibnamefont {Schreiber}},\ and\ \bibinfo {author} {\bibfnamefont
  {M.}~\bibnamefont {Veldhorst}},\ }\href
  {https://doi.org/10.1038/s41534-017-0038-y} {\bibfield  {journal} {\bibinfo
  {journal} {npj Quantum Information}\ }\textbf {\bibinfo {volume} {3}},\
  \bibinfo {pages} {1} (\bibinfo {year} {2017})}\BibitemShut {NoStop}%
\bibitem [{\citenamefont {Veldhorst}\ \emph {et~al.}(2017)\citenamefont
  {Veldhorst}, \citenamefont {Eenink}, \citenamefont {Yang},\ and\
  \citenamefont {Dzurak}}]{Veldhorst:2017p1}%
  \BibitemOpen
  \bibfield  {author} {\bibinfo {author} {\bibfnamefont {M.}~\bibnamefont
  {Veldhorst}}, \bibinfo {author} {\bibfnamefont {H.}~\bibnamefont {Eenink}},
  \bibinfo {author} {\bibfnamefont {C.-H.}\ \bibnamefont {Yang}},\ and\
  \bibinfo {author} {\bibfnamefont {A.~S.}\ \bibnamefont {Dzurak}},\ }\href
  {https://doi.org/10.1038/s41467-017-01905-6} {\bibfield  {journal} {\bibinfo
  {journal} {Nature Communications}\ }\textbf {\bibinfo {volume} {8}},\
  \bibinfo {pages} {1} (\bibinfo {year} {2017})}\BibitemShut {NoStop}%
\bibitem [{\citenamefont {Zajac}\ \emph {et~al.}(2016)\citenamefont {Zajac},
  \citenamefont {Hazard}, \citenamefont {Mi}, \citenamefont {Nielsen},\ and\
  \citenamefont {Petta}}]{Zajac:2016p054013}%
  \BibitemOpen
  \bibfield  {author} {\bibinfo {author} {\bibfnamefont {D.~M.}\ \bibnamefont
  {Zajac}}, \bibinfo {author} {\bibfnamefont {T.~M.}\ \bibnamefont {Hazard}},
  \bibinfo {author} {\bibfnamefont {X.}~\bibnamefont {Mi}}, \bibinfo {author}
  {\bibfnamefont {E.}~\bibnamefont {Nielsen}},\ and\ \bibinfo {author}
  {\bibfnamefont {J.~R.}\ \bibnamefont {Petta}},\ }\href
  {https://doi.org/10.1103/PhysRevApplied.6.054013} {\bibfield  {journal}
  {\bibinfo  {journal} {Phys. Rev. Appl.}\ }\textbf {\bibinfo {volume} {6}},\
  \bibinfo {pages} {054013} (\bibinfo {year} {2016})}\BibitemShut {NoStop}%
\bibitem [{\citenamefont {Zajac}\ \emph {et~al.}(2018)\citenamefont {Zajac},
  \citenamefont {Sigillito}, \citenamefont {Russ}, \citenamefont {Borjans},
  \citenamefont {Taylor}, \citenamefont {Burkard},\ and\ \citenamefont
  {Petta}}]{Zajac:2018p439}%
  \BibitemOpen
  \bibfield  {author} {\bibinfo {author} {\bibfnamefont {D.~M.}\ \bibnamefont
  {Zajac}}, \bibinfo {author} {\bibfnamefont {A.~J.}\ \bibnamefont
  {Sigillito}}, \bibinfo {author} {\bibfnamefont {M.}~\bibnamefont {Russ}},
  \bibinfo {author} {\bibfnamefont {F.}~\bibnamefont {Borjans}}, \bibinfo
  {author} {\bibfnamefont {J.~M.}\ \bibnamefont {Taylor}}, \bibinfo {author}
  {\bibfnamefont {G.}~\bibnamefont {Burkard}},\ and\ \bibinfo {author}
  {\bibfnamefont {J.~R.}\ \bibnamefont {Petta}},\ }\href
  {https://doi.org/10.1126/science.aao5965} {\bibfield  {journal} {\bibinfo
  {journal} {Science}\ }\textbf {\bibinfo {volume} {359}},\ \bibinfo {pages}
  {439} (\bibinfo {year} {2018})}\BibitemShut {NoStop}%
\bibitem [{\citenamefont {Mi}\ \emph {et~al.}(2018)\citenamefont {Mi},
  \citenamefont {Benito}, \citenamefont {Putz}, \citenamefont {Zajac},
  \citenamefont {Taylor}, \citenamefont {Burkard},\ and\ \citenamefont
  {Petta}}]{Mi:2018p599}%
  \BibitemOpen
  \bibfield  {author} {\bibinfo {author} {\bibfnamefont {X.}~\bibnamefont
  {Mi}}, \bibinfo {author} {\bibfnamefont {M.}~\bibnamefont {Benito}}, \bibinfo
  {author} {\bibfnamefont {S.}~\bibnamefont {Putz}}, \bibinfo {author}
  {\bibfnamefont {D.~M.}\ \bibnamefont {Zajac}}, \bibinfo {author}
  {\bibfnamefont {J.~M.}\ \bibnamefont {Taylor}}, \bibinfo {author}
  {\bibfnamefont {G.}~\bibnamefont {Burkard}},\ and\ \bibinfo {author}
  {\bibfnamefont {J.~R.}\ \bibnamefont {Petta}},\ }\href@noop {} {\bibfield
  {journal} {\bibinfo  {journal} {Nature}\ }\textbf {\bibinfo {volume} {555}},\
  \bibinfo {pages} {599} (\bibinfo {year} {2018})}\BibitemShut {NoStop}%
\bibitem [{\citenamefont {Neyens}\ \emph {et~al.}(2019)\citenamefont {Neyens},
  \citenamefont {MacQuarrie}, \citenamefont {Dodson}, \citenamefont {Corrigan},
  \citenamefont {Holman}, \citenamefont {Thorgrimsson}, \citenamefont {Palma},
  \citenamefont {McJunkin}, \citenamefont {Edge}, \citenamefont {Friesen},
  \citenamefont {Coppersmith},\ and\ \citenamefont
  {Eriksson}}]{Neyens:2019p08216}%
  \BibitemOpen
  \bibfield  {author} {\bibinfo {author} {\bibfnamefont {S.~F.}\ \bibnamefont
  {Neyens}}, \bibinfo {author} {\bibfnamefont {E.}~\bibnamefont {MacQuarrie}},
  \bibinfo {author} {\bibfnamefont {J.}~\bibnamefont {Dodson}}, \bibinfo
  {author} {\bibfnamefont {J.}~\bibnamefont {Corrigan}}, \bibinfo {author}
  {\bibfnamefont {N.}~\bibnamefont {Holman}}, \bibinfo {author} {\bibfnamefont
  {B.}~\bibnamefont {Thorgrimsson}}, \bibinfo {author} {\bibfnamefont
  {M.}~\bibnamefont {Palma}}, \bibinfo {author} {\bibfnamefont
  {T.}~\bibnamefont {McJunkin}}, \bibinfo {author} {\bibfnamefont
  {L.}~\bibnamefont {Edge}}, \bibinfo {author} {\bibfnamefont {M.}~\bibnamefont
  {Friesen}}, \bibinfo {author} {\bibfnamefont {S.}~\bibnamefont
  {Coppersmith}},\ and\ \bibinfo {author} {\bibfnamefont {M.}~\bibnamefont
  {Eriksson}},\ }\href {https://doi.org/10.1103/PhysRevApplied.12.064049}
  {\bibfield  {journal} {\bibinfo  {journal} {Phys. Rev. Applied}\ }\textbf
  {\bibinfo {volume} {12}},\ \bibinfo {pages} {064049} (\bibinfo {year}
  {2019})}\BibitemShut {NoStop}%
\bibitem [{\citenamefont {Dodson}\ \emph {et~al.}(2020)\citenamefont {Dodson},
  \citenamefont {Holman}, \citenamefont {Thorgrimsson}, \citenamefont {Neyens},
  \citenamefont {MacQuarrie}, \citenamefont {McJunkin}, \citenamefont {Foote},
  \citenamefont {Edge}, \citenamefont {Coppersmith},\ and\ \citenamefont
  {Eriksson}}]{Dodson:2020p505001}%
  \BibitemOpen
  \bibfield  {author} {\bibinfo {author} {\bibfnamefont {J.~P.}\ \bibnamefont
  {Dodson}}, \bibinfo {author} {\bibfnamefont {N.}~\bibnamefont {Holman}},
  \bibinfo {author} {\bibfnamefont {B.}~\bibnamefont {Thorgrimsson}}, \bibinfo
  {author} {\bibfnamefont {S.~F.}\ \bibnamefont {Neyens}}, \bibinfo {author}
  {\bibfnamefont {E.~F.}\ \bibnamefont {MacQuarrie}}, \bibinfo {author}
  {\bibfnamefont {T.}~\bibnamefont {McJunkin}}, \bibinfo {author}
  {\bibfnamefont {R.~H.}\ \bibnamefont {Foote}}, \bibinfo {author}
  {\bibfnamefont {L.~F.}\ \bibnamefont {Edge}}, \bibinfo {author}
  {\bibfnamefont {S.~N.}\ \bibnamefont {Coppersmith}},\ and\ \bibinfo {author}
  {\bibfnamefont {M.~A.}\ \bibnamefont {Eriksson}},\ }\href
  {https://doi.org/doi:10.1088/1361-6528/abb559} {\bibfield  {journal}
  {\bibinfo  {journal} {Nanotechnology}\ }\textbf {\bibinfo {volume} {31}},\
  \bibinfo {pages} {505001} (\bibinfo {year} {2020})}\BibitemShut {NoStop}%
\bibitem [{\citenamefont {Mills}\ \emph {et~al.}(2019)\citenamefont {Mills},
  \citenamefont {Zajac}, \citenamefont {Gullans}, \citenamefont {Schupp},
  \citenamefont {Hazard},\ and\ \citenamefont {Petta}}]{Mills:2019p1}%
  \BibitemOpen
  \bibfield  {author} {\bibinfo {author} {\bibfnamefont {A.}~\bibnamefont
  {Mills}}, \bibinfo {author} {\bibfnamefont {D.}~\bibnamefont {Zajac}},
  \bibinfo {author} {\bibfnamefont {M.}~\bibnamefont {Gullans}}, \bibinfo
  {author} {\bibfnamefont {F.}~\bibnamefont {Schupp}}, \bibinfo {author}
  {\bibfnamefont {T.}~\bibnamefont {Hazard}},\ and\ \bibinfo {author}
  {\bibfnamefont {J.}~\bibnamefont {Petta}},\ }\href@noop {} {\bibfield
  {journal} {\bibinfo  {journal} {Nature Communications}\ }\textbf {\bibinfo
  {volume} {10}},\ \bibinfo {pages} {1} (\bibinfo {year} {2019})}\BibitemShut
  {NoStop}%
\bibitem [{\citenamefont {Andrews}\ \emph {et~al.}(2019)\citenamefont
  {Andrews}, \citenamefont {Jones}, \citenamefont {Reed}, \citenamefont
  {Jones}, \citenamefont {Ha}, \citenamefont {Jura}, \citenamefont {Kerckhoff},
  \citenamefont {Levendorf}, \citenamefont {Meenehan}, \citenamefont {Merkel},
  \citenamefont {Smith}, \citenamefont {Sun}, \citenamefont {Weinstein},
  \citenamefont {Rakher}, \citenamefont {Ladd},\ and\ \citenamefont
  {Borselli}}]{Andrews:2019p05004}%
  \BibitemOpen
  \bibfield  {author} {\bibinfo {author} {\bibfnamefont {R.~W.}\ \bibnamefont
  {Andrews}}, \bibinfo {author} {\bibfnamefont {C.}~\bibnamefont {Jones}},
  \bibinfo {author} {\bibfnamefont {M.~D.}\ \bibnamefont {Reed}}, \bibinfo
  {author} {\bibfnamefont {A.~M.}\ \bibnamefont {Jones}}, \bibinfo {author}
  {\bibfnamefont {S.~D.}\ \bibnamefont {Ha}}, \bibinfo {author} {\bibfnamefont
  {M.~P.}\ \bibnamefont {Jura}}, \bibinfo {author} {\bibfnamefont
  {J.}~\bibnamefont {Kerckhoff}}, \bibinfo {author} {\bibfnamefont
  {M.}~\bibnamefont {Levendorf}}, \bibinfo {author} {\bibfnamefont
  {S.}~\bibnamefont {Meenehan}}, \bibinfo {author} {\bibfnamefont {S.~T.}\
  \bibnamefont {Merkel}}, \bibinfo {author} {\bibfnamefont {A.}~\bibnamefont
  {Smith}}, \bibinfo {author} {\bibfnamefont {B.}~\bibnamefont {Sun}}, \bibinfo
  {author} {\bibfnamefont {A.~J.}\ \bibnamefont {Weinstein}}, \bibinfo {author}
  {\bibfnamefont {M.~T.}\ \bibnamefont {Rakher}}, \bibinfo {author}
  {\bibfnamefont {T.~D.}\ \bibnamefont {Ladd}},\ and\ \bibinfo {author}
  {\bibfnamefont {M.~G.}\ \bibnamefont {Borselli}},\ }\href
  {https://doi.org/10.1038/s41565-019-0500-4} {\bibfield  {journal} {\bibinfo
  {journal} {Nature Nanotechnology}\ }\textbf {\bibinfo {volume} {14}},\
  \bibinfo {pages} {747} (\bibinfo {year} {2019})}\BibitemShut {NoStop}%
\bibitem [{\citenamefont {Holman}\ \emph {et~al.}(2020)\citenamefont {Holman},
  \citenamefont {Dodson}, \citenamefont {Edge}, \citenamefont {Coppersmith},
  \citenamefont {Friesen}, \citenamefont {McDermott},\ and\ \citenamefont
  {Eriksson}}]{Holman:2020p083502}%
  \BibitemOpen
  \bibfield  {author} {\bibinfo {author} {\bibfnamefont {N.}~\bibnamefont
  {Holman}}, \bibinfo {author} {\bibfnamefont {J.}~\bibnamefont {Dodson}},
  \bibinfo {author} {\bibfnamefont {L.}~\bibnamefont {Edge}}, \bibinfo {author}
  {\bibfnamefont {S.}~\bibnamefont {Coppersmith}}, \bibinfo {author}
  {\bibfnamefont {M.}~\bibnamefont {Friesen}}, \bibinfo {author} {\bibfnamefont
  {R.}~\bibnamefont {McDermott}},\ and\ \bibinfo {author} {\bibfnamefont
  {M.}~\bibnamefont {Eriksson}},\ }\href
  {https://doi.org/doi:10.1063/5.0016248} {\bibfield  {journal} {\bibinfo
  {journal} {Appl. Phys. Lett.}\ }\textbf {\bibinfo {volume} {117}},\ \bibinfo
  {pages} {083502} (\bibinfo {year} {2020})}\BibitemShut {NoStop}%
\bibitem [{\citenamefont {Takeda}\ \emph {et~al.}(2021)\citenamefont {Takeda},
  \citenamefont {Noiri}, \citenamefont {Nakajima}, \citenamefont {Yoneda},
  \citenamefont {Kobayashi},\ and\ \citenamefont
  {Tarucha}}]{Takeda:2021p41565}%
  \BibitemOpen
  \bibfield  {author} {\bibinfo {author} {\bibfnamefont {K.}~\bibnamefont
  {Takeda}}, \bibinfo {author} {\bibfnamefont {A.}~\bibnamefont {Noiri}},
  \bibinfo {author} {\bibfnamefont {T.}~\bibnamefont {Nakajima}}, \bibinfo
  {author} {\bibfnamefont {J.}~\bibnamefont {Yoneda}}, \bibinfo {author}
  {\bibfnamefont {T.}~\bibnamefont {Kobayashi}},\ and\ \bibinfo {author}
  {\bibfnamefont {S.}~\bibnamefont {Tarucha}},\ }\href
  {https://doi.org/10.1038/s41565-021-00925-0} {\bibfield  {journal} {\bibinfo
  {journal} {Nature Nanotechnology}\ }\textbf {\bibinfo {volume} {16}},\
  \bibinfo {pages} {965} (\bibinfo {year} {2021})}\BibitemShut {NoStop}%
\bibitem [{\citenamefont {Ando}\ \emph {et~al.}(1982)\citenamefont {Ando},
  \citenamefont {Fowler},\ and\ \citenamefont {Stern}}]{Ando:1982p437}%
  \BibitemOpen
  \bibfield  {author} {\bibinfo {author} {\bibfnamefont {T.}~\bibnamefont
  {Ando}}, \bibinfo {author} {\bibfnamefont {A.~B.}\ \bibnamefont {Fowler}},\
  and\ \bibinfo {author} {\bibfnamefont {F.}~\bibnamefont {Stern}},\ }\href
  {https://doi.org/10.1103/RevModPhys.54.437} {\bibfield  {journal} {\bibinfo
  {journal} {Rev. Mod. Phys.}\ }\textbf {\bibinfo {volume} {54}},\ \bibinfo
  {pages} {437} (\bibinfo {year} {1982})}\BibitemShut {NoStop}%
\bibitem [{\citenamefont {Boykin}\ \emph
  {et~al.}(2004{\natexlab{a}})\citenamefont {Boykin}, \citenamefont {Klimeck},
  \citenamefont {Eriksson}, \citenamefont {Friesen}, \citenamefont
  {Coppersmith}, \citenamefont {von Allmen}, \citenamefont {Oyafuso},\ and\
  \citenamefont {Lee}}]{Boykin:2004p115}%
  \BibitemOpen
  \bibfield  {author} {\bibinfo {author} {\bibfnamefont {T.~B.}\ \bibnamefont
  {Boykin}}, \bibinfo {author} {\bibfnamefont {G.}~\bibnamefont {Klimeck}},
  \bibinfo {author} {\bibfnamefont {M.~A.}\ \bibnamefont {Eriksson}}, \bibinfo
  {author} {\bibfnamefont {M.}~\bibnamefont {Friesen}}, \bibinfo {author}
  {\bibfnamefont {S.~N.}\ \bibnamefont {Coppersmith}}, \bibinfo {author}
  {\bibfnamefont {P.}~\bibnamefont {von Allmen}}, \bibinfo {author}
  {\bibfnamefont {F.}~\bibnamefont {Oyafuso}},\ and\ \bibinfo {author}
  {\bibfnamefont {S.}~\bibnamefont {Lee}},\ }\href
  {https://doi.org/10.1063/1.1637718} {\bibfield  {journal} {\bibinfo
  {journal} {Appl. Phys. Lett.}\ }\textbf {\bibinfo {volume} {84}},\ \bibinfo
  {pages} {115} (\bibinfo {year} {2004}{\natexlab{a}})}\BibitemShut {NoStop}%
\bibitem [{\citenamefont {Friesen}\ \emph {et~al.}(2006)\citenamefont
  {Friesen}, \citenamefont {Eriksson},\ and\ \citenamefont
  {Coppersmith}}]{Friesen:2006p202106}%
  \BibitemOpen
  \bibfield  {author} {\bibinfo {author} {\bibfnamefont {M.}~\bibnamefont
  {Friesen}}, \bibinfo {author} {\bibfnamefont {M.~A.}\ \bibnamefont
  {Eriksson}},\ and\ \bibinfo {author} {\bibfnamefont {S.~N.}\ \bibnamefont
  {Coppersmith}},\ }\href {https://doi.org/10.1063/1.2387975} {\bibfield
  {journal} {\bibinfo  {journal} {Appl. Phys. Lett.}\ }\textbf {\bibinfo
  {volume} {89}},\ \bibinfo {pages} {202106} (\bibinfo {year}
  {2006})}\BibitemShut {NoStop}%
\bibitem [{\citenamefont {Kharche}\ \emph {et~al.}(2007)\citenamefont
  {Kharche}, \citenamefont {Prada}, \citenamefont {Boykin},\ and\ \citenamefont
  {Klimeck}}]{Kharche:2007p092109}%
  \BibitemOpen
  \bibfield  {author} {\bibinfo {author} {\bibfnamefont {N.}~\bibnamefont
  {Kharche}}, \bibinfo {author} {\bibfnamefont {M.}~\bibnamefont {Prada}},
  \bibinfo {author} {\bibfnamefont {T.~B.}\ \bibnamefont {Boykin}},\ and\
  \bibinfo {author} {\bibfnamefont {G.}~\bibnamefont {Klimeck}},\ }\href
  {https://doi.org/10.1063/1.2591432} {\bibfield  {journal} {\bibinfo
  {journal} {Appl. Phys. Lett.}\ }\textbf {\bibinfo {volume} {90}},\ \bibinfo
  {pages} {092109} (\bibinfo {year} {2007})}\BibitemShut {NoStop}%
\bibitem [{\citenamefont {Culcer}\ \emph {et~al.}(2010)\citenamefont {Culcer},
  \citenamefont {Hu},\ and\ \citenamefont {Das~Sarma}}]{Culcer:2010p205315}%
  \BibitemOpen
  \bibfield  {author} {\bibinfo {author} {\bibfnamefont {D.}~\bibnamefont
  {Culcer}}, \bibinfo {author} {\bibfnamefont {X.}~\bibnamefont {Hu}},\ and\
  \bibinfo {author} {\bibfnamefont {S.}~\bibnamefont {Das~Sarma}},\ }\href
  {https://doi.org/10.1103/PhysRevB.82.205315} {\bibfield  {journal} {\bibinfo
  {journal} {Phys. Rev. B}\ }\textbf {\bibinfo {volume} {82}},\ \bibinfo
  {pages} {205315} (\bibinfo {year} {2010})}\BibitemShut {NoStop}%
\bibitem [{\citenamefont {Gamble}\ \emph {et~al.}(2013)\citenamefont {Gamble},
  \citenamefont {Eriksson}, \citenamefont {Coppersmith},\ and\ \citenamefont
  {Friesen}}]{Gamble:2013p035310}%
  \BibitemOpen
  \bibfield  {author} {\bibinfo {author} {\bibfnamefont {J.~K.}\ \bibnamefont
  {Gamble}}, \bibinfo {author} {\bibfnamefont {M.~A.}\ \bibnamefont
  {Eriksson}}, \bibinfo {author} {\bibfnamefont {S.~N.}\ \bibnamefont
  {Coppersmith}},\ and\ \bibinfo {author} {\bibfnamefont {M.}~\bibnamefont
  {Friesen}},\ }\href@noop {} {\bibfield  {journal} {\bibinfo  {journal} {Phys.
  Rev. B}\ }\textbf {\bibinfo {volume} {88}},\ \bibinfo {pages} {035310}
  (\bibinfo {year} {2013})}\BibitemShut {NoStop}%
\bibitem [{\citenamefont {Boross}\ \emph {et~al.}(2016)\citenamefont {Boross},
  \citenamefont {Sz{\'e}chenyi}, \citenamefont {Culcer},\ and\ \citenamefont
  {P{\'a}lyi}}]{Boross:2016p035438}%
  \BibitemOpen
  \bibfield  {author} {\bibinfo {author} {\bibfnamefont {P.}~\bibnamefont
  {Boross}}, \bibinfo {author} {\bibfnamefont {G.}~\bibnamefont
  {Sz{\'e}chenyi}}, \bibinfo {author} {\bibfnamefont {D.}~\bibnamefont
  {Culcer}},\ and\ \bibinfo {author} {\bibfnamefont {A.}~\bibnamefont
  {P{\'a}lyi}},\ }\href@noop {} {\bibfield  {journal} {\bibinfo  {journal}
  {Physical Review B}\ }\textbf {\bibinfo {volume} {94}},\ \bibinfo {pages}
  {035438} (\bibinfo {year} {2016})}\BibitemShut {NoStop}%
\bibitem [{\citenamefont {Hosseinkhani}\ and\ \citenamefont
  {Burkard}(2020)}]{Hosseinkhani:2020p043180}%
  \BibitemOpen
  \bibfield  {author} {\bibinfo {author} {\bibfnamefont {A.}~\bibnamefont
  {Hosseinkhani}}\ and\ \bibinfo {author} {\bibfnamefont {G.}~\bibnamefont
  {Burkard}},\ }\href {https://doi.org/10.1103/PhysRevResearch.2.043180}
  {\bibfield  {journal} {\bibinfo  {journal} {Phys. Rev. Research}\ }\textbf
  {\bibinfo {volume} {2}},\ \bibinfo {pages} {043180} (\bibinfo {year}
  {2020})}\BibitemShut {NoStop}%
\bibitem [{\citenamefont {Shaji}\ \emph {et~al.}(2008)\citenamefont {Shaji},
  \citenamefont {Simmons}, \citenamefont {Thalakulam}, \citenamefont {Klein},
  \citenamefont {Qin}, \citenamefont {Luo}, \citenamefont {Savage},
  \citenamefont {Lagally}, \citenamefont {Rimberg}, \citenamefont {Joynt},
  \citenamefont {Friesen}, \citenamefont {Blick}, \citenamefont {Coppersmith},\
  and\ \citenamefont {Eriksson}}]{Shaji:2008p540}%
  \BibitemOpen
  \bibfield  {author} {\bibinfo {author} {\bibfnamefont {N.}~\bibnamefont
  {Shaji}}, \bibinfo {author} {\bibfnamefont {C.~B.}\ \bibnamefont {Simmons}},
  \bibinfo {author} {\bibfnamefont {M.}~\bibnamefont {Thalakulam}}, \bibinfo
  {author} {\bibfnamefont {L.~J.}\ \bibnamefont {Klein}}, \bibinfo {author}
  {\bibfnamefont {H.}~\bibnamefont {Qin}}, \bibinfo {author} {\bibfnamefont
  {H.}~\bibnamefont {Luo}}, \bibinfo {author} {\bibfnamefont {D.~E.}\
  \bibnamefont {Savage}}, \bibinfo {author} {\bibfnamefont {M.~G.}\
  \bibnamefont {Lagally}}, \bibinfo {author} {\bibfnamefont {A.~J.}\
  \bibnamefont {Rimberg}}, \bibinfo {author} {\bibfnamefont {R.}~\bibnamefont
  {Joynt}}, \bibinfo {author} {\bibfnamefont {M.}~\bibnamefont {Friesen}},
  \bibinfo {author} {\bibfnamefont {R.~H.}\ \bibnamefont {Blick}}, \bibinfo
  {author} {\bibfnamefont {S.~N.}\ \bibnamefont {Coppersmith}},\ and\ \bibinfo
  {author} {\bibfnamefont {M.~A.}\ \bibnamefont {Eriksson}},\ }\href
  {https://doi.org/doi:10.1038/nphys988} {\bibfield  {journal} {\bibinfo
  {journal} {Nat. Phys.}\ }\textbf {\bibinfo {volume} {4}},\ \bibinfo {pages}
  {540} (\bibinfo {year} {2008})}\BibitemShut {NoStop}%
\bibitem [{\citenamefont {Simmons}\ \emph {et~al.}(2010)\citenamefont
  {Simmons}, \citenamefont {Koh}, \citenamefont {Shaji}, \citenamefont
  {Thalakulam}, \citenamefont {Klein}, \citenamefont {Qin}, \citenamefont
  {Luo}, \citenamefont {Savage}, \citenamefont {Lagally}, \citenamefont
  {Rimberg}, \citenamefont {Joynt}, \citenamefont {Blick}, \citenamefont
  {Friesen}, \citenamefont {Coppersmith},\ and\ \citenamefont
  {Eriksson}}]{Simmons:2010p245312}%
  \BibitemOpen
  \bibfield  {author} {\bibinfo {author} {\bibfnamefont {C.~B.}\ \bibnamefont
  {Simmons}}, \bibinfo {author} {\bibfnamefont {T.~S.}\ \bibnamefont {Koh}},
  \bibinfo {author} {\bibfnamefont {N.}~\bibnamefont {Shaji}}, \bibinfo
  {author} {\bibfnamefont {M.}~\bibnamefont {Thalakulam}}, \bibinfo {author}
  {\bibfnamefont {L.~J.}\ \bibnamefont {Klein}}, \bibinfo {author}
  {\bibfnamefont {H.}~\bibnamefont {Qin}}, \bibinfo {author} {\bibfnamefont
  {H.}~\bibnamefont {Luo}}, \bibinfo {author} {\bibfnamefont {D.~E.}\
  \bibnamefont {Savage}}, \bibinfo {author} {\bibfnamefont {M.~G.}\
  \bibnamefont {Lagally}}, \bibinfo {author} {\bibfnamefont {A.~J.}\
  \bibnamefont {Rimberg}}, \bibinfo {author} {\bibfnamefont {R.}~\bibnamefont
  {Joynt}}, \bibinfo {author} {\bibfnamefont {R.}~\bibnamefont {Blick}},
  \bibinfo {author} {\bibfnamefont {M.}~\bibnamefont {Friesen}}, \bibinfo
  {author} {\bibfnamefont {S.~N.}\ \bibnamefont {Coppersmith}},\ and\ \bibinfo
  {author} {\bibfnamefont {M.~A.}\ \bibnamefont {Eriksson}},\ }\href
  {https://doi.org/10.1103/PhysRevB.82.245312} {\bibfield  {journal} {\bibinfo
  {journal} {Phys. Rev. B}\ }\textbf {\bibinfo {volume} {82}},\ \bibinfo
  {pages} {245312} (\bibinfo {year} {2010})}\BibitemShut {NoStop}%
\bibitem [{\citenamefont {Shi}\ \emph {et~al.}(2011)\citenamefont {Shi},
  \citenamefont {Simmons}, \citenamefont {Prance}, \citenamefont {Gamble},
  \citenamefont {Friesen}, \citenamefont {Savage}, \citenamefont {Lagally},
  \citenamefont {Coppersmith},\ and\ \citenamefont
  {Eriksson}}]{Shi:2011p233108}%
  \BibitemOpen
  \bibfield  {author} {\bibinfo {author} {\bibfnamefont {Z.}~\bibnamefont
  {Shi}}, \bibinfo {author} {\bibfnamefont {C.~B.}\ \bibnamefont {Simmons}},
  \bibinfo {author} {\bibfnamefont {J.}~\bibnamefont {Prance}}, \bibinfo
  {author} {\bibfnamefont {J.~K.}\ \bibnamefont {Gamble}}, \bibinfo {author}
  {\bibfnamefont {M.}~\bibnamefont {Friesen}}, \bibinfo {author} {\bibfnamefont
  {D.~E.}\ \bibnamefont {Savage}}, \bibinfo {author} {\bibfnamefont {M.~G.}\
  \bibnamefont {Lagally}}, \bibinfo {author} {\bibfnamefont {S.~N.}\
  \bibnamefont {Coppersmith}},\ and\ \bibinfo {author} {\bibfnamefont {M.~A.}\
  \bibnamefont {Eriksson}},\ }\href {https://doi.org/10.1063/1.3666232}
  {\bibfield  {journal} {\bibinfo  {journal} {Appl. Phys. Lett.}\ }\textbf
  {\bibinfo {volume} {99}},\ \bibinfo {pages} {233108} (\bibinfo {year}
  {2011})}\BibitemShut {NoStop}%
\bibitem [{\citenamefont {Borselli}\ \emph {et~al.}(2011)\citenamefont
  {Borselli}, \citenamefont {Ross}, \citenamefont {Kiselev}, \citenamefont
  {Croke}, \citenamefont {Holabird}, \citenamefont {Deelman}, \citenamefont
  {Warren}, \citenamefont {Alvarado-Rodriguez}, \citenamefont {Milosavljevic},
  \citenamefont {Ku}, \citenamefont {Wong}, \citenamefont {Schmitz},
  \citenamefont {Sokolich}, \citenamefont {Gyure},\ and\ \citenamefont
  {Hunter}}]{Borselli:2011p123118}%
  \BibitemOpen
  \bibfield  {author} {\bibinfo {author} {\bibfnamefont {M.~G.}\ \bibnamefont
  {Borselli}}, \bibinfo {author} {\bibfnamefont {R.~S.}\ \bibnamefont {Ross}},
  \bibinfo {author} {\bibfnamefont {A.~A.}\ \bibnamefont {Kiselev}}, \bibinfo
  {author} {\bibfnamefont {E.~T.}\ \bibnamefont {Croke}}, \bibinfo {author}
  {\bibfnamefont {K.~S.}\ \bibnamefont {Holabird}}, \bibinfo {author}
  {\bibfnamefont {P.~W.}\ \bibnamefont {Deelman}}, \bibinfo {author}
  {\bibfnamefont {L.~D.}\ \bibnamefont {Warren}}, \bibinfo {author}
  {\bibfnamefont {I.}~\bibnamefont {Alvarado-Rodriguez}}, \bibinfo {author}
  {\bibfnamefont {I.}~\bibnamefont {Milosavljevic}}, \bibinfo {author}
  {\bibfnamefont {F.~C.}\ \bibnamefont {Ku}}, \bibinfo {author} {\bibfnamefont
  {W.~S.}\ \bibnamefont {Wong}}, \bibinfo {author} {\bibfnamefont {A.~E.}\
  \bibnamefont {Schmitz}}, \bibinfo {author} {\bibfnamefont {M.}~\bibnamefont
  {Sokolich}}, \bibinfo {author} {\bibfnamefont {M.~F.}\ \bibnamefont
  {Gyure}},\ and\ \bibinfo {author} {\bibfnamefont {A.~T.}\ \bibnamefont
  {Hunter}},\ }\href {https://doi.org/10.1063/1.3569717} {\bibfield  {journal}
  {\bibinfo  {journal} {Appl. Phys. Lett.}\ }\textbf {\bibinfo {volume} {98}},\
  \bibinfo {pages} {123118} (\bibinfo {year} {2011})}\BibitemShut {NoStop}%
\bibitem [{\citenamefont {Kim}\ \emph {et~al.}(2014)\citenamefont {Kim},
  \citenamefont {Shi}, \citenamefont {Simmons}, \citenamefont {Ward},
  \citenamefont {Prance}, \citenamefont {Koh}, \citenamefont {Gamble},
  \citenamefont {Savage}, \citenamefont {Lagally}, \citenamefont {Friesen},
  \citenamefont {Coppersmith},\ and\ \citenamefont {Eriksson}}]{Kim:2014p70}%
  \BibitemOpen
  \bibfield  {author} {\bibinfo {author} {\bibfnamefont {D.}~\bibnamefont
  {Kim}}, \bibinfo {author} {\bibfnamefont {Z.}~\bibnamefont {Shi}}, \bibinfo
  {author} {\bibfnamefont {C.~B.}\ \bibnamefont {Simmons}}, \bibinfo {author}
  {\bibfnamefont {D.~R.}\ \bibnamefont {Ward}}, \bibinfo {author}
  {\bibfnamefont {J.~R.}\ \bibnamefont {Prance}}, \bibinfo {author}
  {\bibfnamefont {T.~S.}\ \bibnamefont {Koh}}, \bibinfo {author} {\bibfnamefont
  {J.~K.}\ \bibnamefont {Gamble}}, \bibinfo {author} {\bibfnamefont {D.~E.}\
  \bibnamefont {Savage}}, \bibinfo {author} {\bibfnamefont {M.~G.}\
  \bibnamefont {Lagally}}, \bibinfo {author} {\bibfnamefont {M.}~\bibnamefont
  {Friesen}}, \bibinfo {author} {\bibfnamefont {S.~N.}\ \bibnamefont
  {Coppersmith}},\ and\ \bibinfo {author} {\bibfnamefont {M.~A.}\ \bibnamefont
  {Eriksson}},\ }\href@noop {} {\bibfield  {journal} {\bibinfo  {journal}
  {Nature}\ }\textbf {\bibinfo {volume} {511}},\ \bibinfo {pages} {70}
  (\bibinfo {year} {2014})}\BibitemShut {NoStop}%
\bibitem [{\citenamefont {Schoenfield}\ \emph {et~al.}(2017)\citenamefont
  {Schoenfield}, \citenamefont {Freeman},\ and\ \citenamefont
  {Jiang}}]{Schoenfield:2017p64}%
  \BibitemOpen
  \bibfield  {author} {\bibinfo {author} {\bibfnamefont {J.~S.}\ \bibnamefont
  {Schoenfield}}, \bibinfo {author} {\bibfnamefont {B.~M.}\ \bibnamefont
  {Freeman}},\ and\ \bibinfo {author} {\bibfnamefont {H.}~\bibnamefont
  {Jiang}},\ }\href {https://doi.org/10.1038/s41467-017-00073-x} {\bibfield
  {journal} {\bibinfo  {journal} {Nature Commun.}\ }\textbf {\bibinfo {volume}
  {8}},\ \bibinfo {pages} {64} (\bibinfo {year} {2017})}\BibitemShut {NoStop}%
\bibitem [{\citenamefont {Mi}\ \emph {et~al.}(2017)\citenamefont {Mi},
  \citenamefont {P\'eterfalvi}, \citenamefont {Burkard},\ and\ \citenamefont
  {Petta}}]{Mi:2018p76803}%
  \BibitemOpen
  \bibfield  {author} {\bibinfo {author} {\bibfnamefont {X.}~\bibnamefont
  {Mi}}, \bibinfo {author} {\bibfnamefont {C.~G.}\ \bibnamefont
  {P\'eterfalvi}}, \bibinfo {author} {\bibfnamefont {G.}~\bibnamefont
  {Burkard}},\ and\ \bibinfo {author} {\bibfnamefont {J.~R.}\ \bibnamefont
  {Petta}},\ }\href {https://doi.org/10.1103/PhysRevLett.119.176803} {\bibfield
   {journal} {\bibinfo  {journal} {Phys. Rev. Lett.}\ }\textbf {\bibinfo
  {volume} {119}},\ \bibinfo {pages} {176803} (\bibinfo {year}
  {2017})}\BibitemShut {NoStop}%
\bibitem [{\citenamefont {Jones}\ \emph {et~al.}(2019)\citenamefont {Jones},
  \citenamefont {Pritchett}, \citenamefont {Chen}, \citenamefont {Keating},
  \citenamefont {Andrews}, \citenamefont {Blumoff}, \citenamefont {De~Lorenzo},
  \citenamefont {Eng}, \citenamefont {Ha}, \citenamefont {Kiselev},
  \citenamefont {Meenehan}, \citenamefont {Merkel}, \citenamefont {Wright},
  \citenamefont {Edge}, \citenamefont {RS}, \citenamefont {MT}, \citenamefont
  {MG},\ and\ \citenamefont {Hunter}}]{Jones:2019p014026}%
  \BibitemOpen
  \bibfield  {author} {\bibinfo {author} {\bibfnamefont {A.}~\bibnamefont
  {Jones}}, \bibinfo {author} {\bibfnamefont {E.}~\bibnamefont {Pritchett}},
  \bibinfo {author} {\bibfnamefont {E.}~\bibnamefont {Chen}}, \bibinfo {author}
  {\bibfnamefont {T.}~\bibnamefont {Keating}}, \bibinfo {author} {\bibfnamefont
  {R.}~\bibnamefont {Andrews}}, \bibinfo {author} {\bibfnamefont
  {J.}~\bibnamefont {Blumoff}}, \bibinfo {author} {\bibfnamefont
  {L.}~\bibnamefont {De~Lorenzo}}, \bibinfo {author} {\bibfnamefont
  {K.}~\bibnamefont {Eng}}, \bibinfo {author} {\bibfnamefont {S.}~\bibnamefont
  {Ha}}, \bibinfo {author} {\bibfnamefont {A.}~\bibnamefont {Kiselev}},
  \bibinfo {author} {\bibfnamefont {S.}~\bibnamefont {Meenehan}}, \bibinfo
  {author} {\bibfnamefont {S.}~\bibnamefont {Merkel}}, \bibinfo {author}
  {\bibfnamefont {J.}~\bibnamefont {Wright}}, \bibinfo {author} {\bibfnamefont
  {L.}~\bibnamefont {Edge}}, \bibinfo {author} {\bibfnamefont {R.}~\bibnamefont
  {RS}}, \bibinfo {author} {\bibfnamefont {R.}~\bibnamefont {MT}}, \bibinfo
  {author} {\bibfnamefont {B.}~\bibnamefont {MG}},\ and\ \bibinfo {author}
  {\bibfnamefont {A.}~\bibnamefont {Hunter}},\ }\href@noop {} {\bibfield
  {journal} {\bibinfo  {journal} {Physical Review Applied}\ }\textbf {\bibinfo
  {volume} {12}},\ \bibinfo {pages} {014026} (\bibinfo {year}
  {2019})}\BibitemShut {NoStop}%
\bibitem [{\citenamefont {Hollmann}\ \emph {et~al.}(2020)\citenamefont
  {Hollmann}, \citenamefont {Struck}, \citenamefont {Langrock}, \citenamefont
  {Schmidbauer}, \citenamefont {Schauer}, \citenamefont {Leonhardt},
  \citenamefont {Sawano}, \citenamefont {Riemann}, \citenamefont {Abrosimov},
  \citenamefont {Bougeard},\ and\ \citenamefont
  {Schreiber}}]{Hollmann:2020p34068}%
  \BibitemOpen
  \bibfield  {author} {\bibinfo {author} {\bibfnamefont {A.}~\bibnamefont
  {Hollmann}}, \bibinfo {author} {\bibfnamefont {T.}~\bibnamefont {Struck}},
  \bibinfo {author} {\bibfnamefont {V.}~\bibnamefont {Langrock}}, \bibinfo
  {author} {\bibfnamefont {A.}~\bibnamefont {Schmidbauer}}, \bibinfo {author}
  {\bibfnamefont {F.}~\bibnamefont {Schauer}}, \bibinfo {author} {\bibfnamefont
  {T.}~\bibnamefont {Leonhardt}}, \bibinfo {author} {\bibfnamefont
  {K.}~\bibnamefont {Sawano}}, \bibinfo {author} {\bibfnamefont
  {H.}~\bibnamefont {Riemann}}, \bibinfo {author} {\bibfnamefont {N.~V.}\
  \bibnamefont {Abrosimov}}, \bibinfo {author} {\bibfnamefont {D.}~\bibnamefont
  {Bougeard}},\ and\ \bibinfo {author} {\bibfnamefont {L.~R.}\ \bibnamefont
  {Schreiber}},\ }\href {https://doi.org/10.1103/PhysRevApplied.13.034068}
  {\bibfield  {journal} {\bibinfo  {journal} {Phys. Rev. Applied}\ }\textbf
  {\bibinfo {volume} {13}},\ \bibinfo {pages} {034068} (\bibinfo {year}
  {2020})}\BibitemShut {NoStop}%
\bibitem [{\citenamefont {Chen}\ \emph {et~al.}(2021)\citenamefont {Chen},
  \citenamefont {Raach}, \citenamefont {Pan}, \citenamefont {Kiselev},
  \citenamefont {Acuna}, \citenamefont {Blumoff}, \citenamefont {Brecht},
  \citenamefont {Choi}, \citenamefont {Ha}, \citenamefont {Hulbert},
  \citenamefont {Jura}, \citenamefont {Keating}, \citenamefont {Noah},
  \citenamefont {Sun}, \citenamefont {Thomas}, \citenamefont {Borselli},
  \citenamefont {Jackson}, \citenamefont {Rakher},\ and\ \citenamefont
  {Ross}}]{Chen:2020p44033}%
  \BibitemOpen
  \bibfield  {author} {\bibinfo {author} {\bibfnamefont {E.~H.}\ \bibnamefont
  {Chen}}, \bibinfo {author} {\bibfnamefont {K.}~\bibnamefont {Raach}},
  \bibinfo {author} {\bibfnamefont {A.}~\bibnamefont {Pan}}, \bibinfo {author}
  {\bibfnamefont {A.~A.}\ \bibnamefont {Kiselev}}, \bibinfo {author}
  {\bibfnamefont {E.}~\bibnamefont {Acuna}}, \bibinfo {author} {\bibfnamefont
  {J.~Z.}\ \bibnamefont {Blumoff}}, \bibinfo {author} {\bibfnamefont
  {T.}~\bibnamefont {Brecht}}, \bibinfo {author} {\bibfnamefont {M.~D.}\
  \bibnamefont {Choi}}, \bibinfo {author} {\bibfnamefont {W.}~\bibnamefont
  {Ha}}, \bibinfo {author} {\bibfnamefont {D.~R.}\ \bibnamefont {Hulbert}},
  \bibinfo {author} {\bibfnamefont {M.~P.}\ \bibnamefont {Jura}}, \bibinfo
  {author} {\bibfnamefont {T.~E.}\ \bibnamefont {Keating}}, \bibinfo {author}
  {\bibfnamefont {R.}~\bibnamefont {Noah}}, \bibinfo {author} {\bibfnamefont
  {B.}~\bibnamefont {Sun}}, \bibinfo {author} {\bibfnamefont {B.~J.}\
  \bibnamefont {Thomas}}, \bibinfo {author} {\bibfnamefont {M.~G.}\
  \bibnamefont {Borselli}}, \bibinfo {author} {\bibfnamefont {C.}~\bibnamefont
  {Jackson}}, \bibinfo {author} {\bibfnamefont {M.~T.}\ \bibnamefont
  {Rakher}},\ and\ \bibinfo {author} {\bibfnamefont {R.~S.}\ \bibnamefont
  {Ross}},\ }\href {https://doi.org/10.1103/PhysRevApplied.15.044033}
  {\bibfield  {journal} {\bibinfo  {journal} {Phys. Rev. Applied}\ }\textbf
  {\bibinfo {volume} {15}},\ \bibinfo {pages} {044033} (\bibinfo {year}
  {2021})}\BibitemShut {NoStop}%
\bibitem [{\citenamefont {Szabo}(1996)}]{Szabo:1996p91861}%
  \BibitemOpen
  \bibfield  {author} {\bibinfo {author} {\bibfnamefont {A.}~\bibnamefont
  {Szabo}},\ }\href@noop {} {\emph {\bibinfo {title} {Modern Quantum Chemistry:
  Introduction to Advanced Electronic Structure Theory (Dover Books on
  Chemistry)}}}\ (\bibinfo  {publisher} {Dover Publications},\ \bibinfo {year}
  {1996})\BibitemShut {NoStop}%
\bibitem [{\citenamefont {Boykin}\ \emph
  {et~al.}(2004{\natexlab{b}})\citenamefont {Boykin}, \citenamefont {Klimeck},
  \citenamefont {Friesen}, \citenamefont {Coppersmith}, \citenamefont
  {vonAllmen}, \citenamefont {Oyafuso},\ and\ \citenamefont
  {Lee}}]{Boykin:2004p165325}%
  \BibitemOpen
  \bibfield  {author} {\bibinfo {author} {\bibfnamefont {T.~B.}\ \bibnamefont
  {Boykin}}, \bibinfo {author} {\bibfnamefont {G.}~\bibnamefont {Klimeck}},
  \bibinfo {author} {\bibfnamefont {M.}~\bibnamefont {Friesen}}, \bibinfo
  {author} {\bibfnamefont {S.~N.}\ \bibnamefont {Coppersmith}}, \bibinfo
  {author} {\bibfnamefont {P.}~\bibnamefont {vonAllmen}}, \bibinfo {author}
  {\bibfnamefont {F.}~\bibnamefont {Oyafuso}},\ and\ \bibinfo {author}
  {\bibfnamefont {S.}~\bibnamefont {Lee}},\ }\href@noop {} {\bibfield
  {journal} {\bibinfo  {journal} {Phys. Rev. B}\ }\textbf {\bibinfo {volume}
  {70}},\ \bibinfo {pages} {165325} (\bibinfo {year}
  {2004}{\natexlab{b}})}\BibitemShut {NoStop}%
\bibitem [{\citenamefont {Elzerman}\ \emph {et~al.}(2004)\citenamefont
  {Elzerman}, \citenamefont {Hanson}, \citenamefont {Willems~van Beveren},
  \citenamefont {Vandersypen},\ and\ \citenamefont
  {Kouwenhoven}}]{Elzerman:2004p731}%
  \BibitemOpen
  \bibfield  {author} {\bibinfo {author} {\bibfnamefont {J.~M.}\ \bibnamefont
  {Elzerman}}, \bibinfo {author} {\bibfnamefont {R.}~\bibnamefont {Hanson}},
  \bibinfo {author} {\bibfnamefont {L.~H.}\ \bibnamefont {Willems~van
  Beveren}}, \bibinfo {author} {\bibfnamefont {L.~M.~K.}\ \bibnamefont
  {Vandersypen}},\ and\ \bibinfo {author} {\bibfnamefont {L.~P.}\ \bibnamefont
  {Kouwenhoven}},\ }\href {https://doi.org/10.1063/1.1757023} {\bibfield
  {journal} {\bibinfo  {journal} {Appl. Phys. Lett.}\ }\textbf {\bibinfo
  {volume} {84}},\ \bibinfo {pages} {4617} (\bibinfo {year}
  {2004})}\BibitemShut {NoStop}%
\bibitem [{\citenamefont {Yang}\ \emph {et~al.}(2012)\citenamefont {Yang},
  \citenamefont {Lim}, \citenamefont {Lai}, \citenamefont {Rossi},
  \citenamefont {Morello},\ and\ \citenamefont {Dzurak}}]{Yang:2012p15319}%
  \BibitemOpen
  \bibfield  {author} {\bibinfo {author} {\bibfnamefont {C.~H.}\ \bibnamefont
  {Yang}}, \bibinfo {author} {\bibfnamefont {W.~H.}\ \bibnamefont {Lim}},
  \bibinfo {author} {\bibfnamefont {N.~S.}\ \bibnamefont {Lai}}, \bibinfo
  {author} {\bibfnamefont {A.}~\bibnamefont {Rossi}}, \bibinfo {author}
  {\bibfnamefont {A.}~\bibnamefont {Morello}},\ and\ \bibinfo {author}
  {\bibfnamefont {A.~S.}\ \bibnamefont {Dzurak}},\ }\href
  {https://doi.org/10.1103/PhysRevB.86.115319} {\bibfield  {journal} {\bibinfo
  {journal} {Phys. Rev. B}\ }\textbf {\bibinfo {volume} {86}},\ \bibinfo
  {pages} {115319} (\bibinfo {year} {2012})}\BibitemShut {NoStop}%
\bibitem [{\citenamefont {Gamble}\ \emph {et~al.}(2016)\citenamefont {Gamble},
  \citenamefont {Harvey-Collard}, \citenamefont {Jacobson}, \citenamefont
  {Baczewski}, \citenamefont {Nielsen}, \citenamefont {Maurer}, \citenamefont
  {Monta{\~n}o}, \citenamefont {Rudolph}, \citenamefont {Carroll},
  \citenamefont {Yang}, \citenamefont {Rossi}, \citenamefont {Dzurak},\ and\
  \citenamefont {Muller}}]{Gamble:2016p253101}%
  \BibitemOpen
  \bibfield  {author} {\bibinfo {author} {\bibfnamefont {J.~K.}\ \bibnamefont
  {Gamble}}, \bibinfo {author} {\bibfnamefont {P.}~\bibnamefont
  {Harvey-Collard}}, \bibinfo {author} {\bibfnamefont {N.~T.}\ \bibnamefont
  {Jacobson}}, \bibinfo {author} {\bibfnamefont {A.~D.}\ \bibnamefont
  {Baczewski}}, \bibinfo {author} {\bibfnamefont {E.}~\bibnamefont {Nielsen}},
  \bibinfo {author} {\bibfnamefont {L.}~\bibnamefont {Maurer}}, \bibinfo
  {author} {\bibfnamefont {I.}~\bibnamefont {Monta{\~n}o}}, \bibinfo {author}
  {\bibfnamefont {M.}~\bibnamefont {Rudolph}}, \bibinfo {author} {\bibfnamefont
  {M.}~\bibnamefont {Carroll}}, \bibinfo {author} {\bibfnamefont
  {C.}~\bibnamefont {Yang}}, \bibinfo {author} {\bibfnamefont {A.}~\bibnamefont
  {Rossi}}, \bibinfo {author} {\bibfnamefont {A.}~\bibnamefont {Dzurak}},\ and\
  \bibinfo {author} {\bibfnamefont {R.~P.}\ \bibnamefont {Muller}},\ }\href
  {https://doi.org/10.1063/1.4972514} {\bibfield  {journal} {\bibinfo
  {journal} {Appl. Phys. Lett.}\ }\textbf {\bibinfo {volume} {109}},\ \bibinfo
  {pages} {253101} (\bibinfo {year} {2016})}\BibitemShut {NoStop}%
\bibitem [{\citenamefont {Field}\ \emph {et~al.}(1993)\citenamefont {Field},
  \citenamefont {Smith}, \citenamefont {Pepper}, \citenamefont {Ritchie},
  \citenamefont {Frost}, \citenamefont {Jones},\ and\ \citenamefont
  {Hasko}}]{Field:1993p1477}%
  \BibitemOpen
  \bibfield  {author} {\bibinfo {author} {\bibfnamefont {M.}~\bibnamefont
  {Field}}, \bibinfo {author} {\bibfnamefont {C.~G.}\ \bibnamefont {Smith}},
  \bibinfo {author} {\bibfnamefont {M.}~\bibnamefont {Pepper}}, \bibinfo
  {author} {\bibfnamefont {D.~A.}\ \bibnamefont {Ritchie}}, \bibinfo {author}
  {\bibfnamefont {J.~E.~F.}\ \bibnamefont {Frost}}, \bibinfo {author}
  {\bibfnamefont {G.~A.~C.}\ \bibnamefont {Jones}},\ and\ \bibinfo {author}
  {\bibfnamefont {D.~G.}\ \bibnamefont {Hasko}},\ }\href
  {https://doi.org/10.1103/PhysRevLett.70.1311} {\bibfield  {journal} {\bibinfo
   {journal} {Phys. Rev. Lett.}\ }\textbf {\bibinfo {volume} {70}},\ \bibinfo
  {pages} {1311} (\bibinfo {year} {1993})}\BibitemShut {NoStop}%
\bibitem [{\citenamefont {Kouwenhoven}\ \emph {et~al.}(1997)\citenamefont
  {Kouwenhoven}, \citenamefont {Marcus}, \citenamefont {McEuen}, \citenamefont
  {Tarucha}, \citenamefont {Westervelt},\ and\ \citenamefont
  {Wingreen}}]{Kouwenhoven:1997p1384}%
  \BibitemOpen
  \bibfield  {author} {\bibinfo {author} {\bibfnamefont {L.~P.}\ \bibnamefont
  {Kouwenhoven}}, \bibinfo {author} {\bibfnamefont {C.~M.}\ \bibnamefont
  {Marcus}}, \bibinfo {author} {\bibfnamefont {P.~L.}\ \bibnamefont {McEuen}},
  \bibinfo {author} {\bibfnamefont {S.}~\bibnamefont {Tarucha}}, \bibinfo
  {author} {\bibfnamefont {R.~M.}\ \bibnamefont {Westervelt}},\ and\ \bibinfo
  {author} {\bibfnamefont {N.~S.}\ \bibnamefont {Wingreen}},\ }\bibinfo {title}
  {Mesoscopic electron transport}\ (\bibinfo  {publisher} {Kluwer},\ \bibinfo
  {year} {1997})\ Chap.\ \bibinfo {chapter} {Electron Transport in Quantum
  Dots}, p.\ \bibinfo {pages} {105}\BibitemShut {NoStop}%
\bibitem [{\citenamefont {Lim}\ \emph {et~al.}(2009)\citenamefont {Lim},
  \citenamefont {Zwanenburg}, \citenamefont {Huebl}, \citenamefont
  {M{\"o}tt{\"o}nen}, \citenamefont {Chan}, \citenamefont {Morello},\ and\
  \citenamefont {Dzurak}}]{Lim:2009p242102}%
  \BibitemOpen
  \bibfield  {author} {\bibinfo {author} {\bibfnamefont {W.~H.}\ \bibnamefont
  {Lim}}, \bibinfo {author} {\bibfnamefont {F.~A.}\ \bibnamefont {Zwanenburg}},
  \bibinfo {author} {\bibfnamefont {H.}~\bibnamefont {Huebl}}, \bibinfo
  {author} {\bibfnamefont {M.}~\bibnamefont {M{\"o}tt{\"o}nen}}, \bibinfo
  {author} {\bibfnamefont {K.~W.}\ \bibnamefont {Chan}}, \bibinfo {author}
  {\bibfnamefont {A.}~\bibnamefont {Morello}},\ and\ \bibinfo {author}
  {\bibfnamefont {A.~S.}\ \bibnamefont {Dzurak}},\ }\href@noop {} {\bibfield
  {journal} {\bibinfo  {journal} {Appl. Phys. Lett.}\ }\textbf {\bibinfo
  {volume} {95}},\ \bibinfo {pages} {242102} (\bibinfo {year}
  {2009})}\BibitemShut {NoStop}%
\bibitem [{\citenamefont {Lim}\ \emph {et~al.}(2011)\citenamefont {Lim},
  \citenamefont {Yang}, \citenamefont {Zwanenburg},\ and\ \citenamefont
  {Dzurak}}]{Lim:2011p35704}%
  \BibitemOpen
  \bibfield  {author} {\bibinfo {author} {\bibfnamefont {W.~H.}\ \bibnamefont
  {Lim}}, \bibinfo {author} {\bibfnamefont {C.~H.}\ \bibnamefont {Yang}},
  \bibinfo {author} {\bibfnamefont {F.~A.}\ \bibnamefont {Zwanenburg}},\ and\
  \bibinfo {author} {\bibfnamefont {A.~S.}\ \bibnamefont {Dzurak}},\ }\href
  {https://doi.org/10.1088/0957-4484/22/33/335704} {\bibfield  {journal}
  {\bibinfo  {journal} {Nanotechnology}\ }\textbf {\bibinfo {volume} {22}},\
  \bibinfo {pages} {335704} (\bibinfo {year} {2011})}\BibitemShut {NoStop}%
\bibitem [{\citenamefont {Friesen}\ and\ \citenamefont
  {Coppersmith}(2010)}]{Friesen:2010p115324}%
  \BibitemOpen
  \bibfield  {author} {\bibinfo {author} {\bibfnamefont {M.}~\bibnamefont
  {Friesen}}\ and\ \bibinfo {author} {\bibfnamefont {S.~N.}\ \bibnamefont
  {Coppersmith}},\ }\href {https://doi.org/10.1103/PhysRevB.81.115324}
  {\bibfield  {journal} {\bibinfo  {journal} {Phys. Rev. B}\ }\textbf {\bibinfo
  {volume} {81}},\ \bibinfo {pages} {115324} (\bibinfo {year}
  {2010})}\BibitemShut {NoStop}%
\bibitem [{\citenamefont {McJunkin}\ \emph {et~al.}(2021)\citenamefont
  {McJunkin}, \citenamefont {MacQuarrie}, \citenamefont {Tom}, \citenamefont
  {Neyens}, \citenamefont {Dodson}, \citenamefont {Thorgrimsson}, \citenamefont
  {Corrigan}, \citenamefont {Ercan}, \citenamefont {Savage}, \citenamefont
  {Lagally}, \citenamefont {Joynt}, \citenamefont {Coppersmith}, \citenamefont
  {Friesen},\ and\ \citenamefont {Eriksson}}]{McJunkin:2021p85406}%
  \BibitemOpen
  \bibfield  {author} {\bibinfo {author} {\bibfnamefont {T.}~\bibnamefont
  {McJunkin}}, \bibinfo {author} {\bibfnamefont {E.~R.}\ \bibnamefont
  {MacQuarrie}}, \bibinfo {author} {\bibfnamefont {L.}~\bibnamefont {Tom}},
  \bibinfo {author} {\bibfnamefont {S.~F.}\ \bibnamefont {Neyens}}, \bibinfo
  {author} {\bibfnamefont {J.~P.}\ \bibnamefont {Dodson}}, \bibinfo {author}
  {\bibfnamefont {B.}~\bibnamefont {Thorgrimsson}}, \bibinfo {author}
  {\bibfnamefont {J.}~\bibnamefont {Corrigan}}, \bibinfo {author}
  {\bibfnamefont {H.~E.}\ \bibnamefont {Ercan}}, \bibinfo {author}
  {\bibfnamefont {D.~E.}\ \bibnamefont {Savage}}, \bibinfo {author}
  {\bibfnamefont {M.~G.}\ \bibnamefont {Lagally}}, \bibinfo {author}
  {\bibfnamefont {R.}~\bibnamefont {Joynt}}, \bibinfo {author} {\bibfnamefont
  {S.~N.}\ \bibnamefont {Coppersmith}}, \bibinfo {author} {\bibfnamefont
  {M.}~\bibnamefont {Friesen}},\ and\ \bibinfo {author} {\bibfnamefont {M.~A.}\
  \bibnamefont {Eriksson}},\ }\href
  {https://doi.org/10.1103/PhysRevB.104.085406} {\bibfield  {journal} {\bibinfo
   {journal} {Phys. Rev. B}\ }\textbf {\bibinfo {volume} {104}},\ \bibinfo
  {pages} {085406} (\bibinfo {year} {2021})}\BibitemShut {NoStop}%
\bibitem [{\citenamefont {Tilka}\ \emph {et~al.}(2016)\citenamefont {Tilka},
  \citenamefont {Park}, \citenamefont {Ahn}, \citenamefont {Pateras},
  \citenamefont {Sampson}, \citenamefont {Savage}, \citenamefont {Prance},
  \citenamefont {Simmons}, \citenamefont {Coppersmith}, \citenamefont
  {Eriksson}, \citenamefont {Lagally}, \citenamefont {Holt},\ and\
  \citenamefont {Evans}}]{Tilka:2016p55043}%
  \BibitemOpen
  \bibfield  {author} {\bibinfo {author} {\bibfnamefont {J.~A.}\ \bibnamefont
  {Tilka}}, \bibinfo {author} {\bibfnamefont {J.}~\bibnamefont {Park}},
  \bibinfo {author} {\bibfnamefont {Y.}~\bibnamefont {Ahn}}, \bibinfo {author}
  {\bibfnamefont {A.}~\bibnamefont {Pateras}}, \bibinfo {author} {\bibfnamefont
  {K.~C.}\ \bibnamefont {Sampson}}, \bibinfo {author} {\bibfnamefont {D.~E.}\
  \bibnamefont {Savage}}, \bibinfo {author} {\bibfnamefont {J.~R.}\
  \bibnamefont {Prance}}, \bibinfo {author} {\bibfnamefont {C.~B.}\
  \bibnamefont {Simmons}}, \bibinfo {author} {\bibfnamefont {S.~N.}\
  \bibnamefont {Coppersmith}}, \bibinfo {author} {\bibfnamefont {M.~A.}\
  \bibnamefont {Eriksson}}, \bibinfo {author} {\bibfnamefont {M.~G.}\
  \bibnamefont {Lagally}}, \bibinfo {author} {\bibfnamefont {M.~V.}\
  \bibnamefont {Holt}},\ and\ \bibinfo {author} {\bibfnamefont {P.~G.}\
  \bibnamefont {Evans}},\ }\href {https://doi.org/10.1063/1.4955043} {\bibfield
   {journal} {\bibinfo  {journal} {Journal of Applied Physics}\ }\textbf
  {\bibinfo {volume} {120}},\ \bibinfo {pages} {015304} (\bibinfo {year}
  {2016})}\BibitemShut {NoStop}%
\bibitem [{\citenamefont {Dyck}\ \emph {et~al.}(2017)\citenamefont {Dyck},
  \citenamefont {Leonard}, \citenamefont {Edge}, \citenamefont {Jackson},
  \citenamefont {Pritchett}, \citenamefont {Deelman},\ and\ \citenamefont
  {Poplawsky}}]{Osti:2017p74844}%
  \BibitemOpen
  \bibfield  {author} {\bibinfo {author} {\bibfnamefont {O.~E.}\ \bibnamefont
  {Dyck}}, \bibinfo {author} {\bibfnamefont {D.~N.}\ \bibnamefont {Leonard}},
  \bibinfo {author} {\bibfnamefont {L.}~\bibnamefont {Edge}}, \bibinfo {author}
  {\bibfnamefont {C.}~\bibnamefont {Jackson}}, \bibinfo {author} {\bibfnamefont
  {E.~J.}\ \bibnamefont {Pritchett}}, \bibinfo {author} {\bibfnamefont {P.~W.}\
  \bibnamefont {Deelman}},\ and\ \bibinfo {author} {\bibfnamefont {J.~D.}\
  \bibnamefont {Poplawsky}},\ }\bibfield  {journal} {\bibinfo  {journal}
  {Advanced Materials Interfaces}\ }\textbf {\bibinfo {volume} {4}},\ \href
  {https://doi.org/10.1002/admi.201700622} {10.1002/admi.201700622} (\bibinfo
  {year} {2017})\BibitemShut {NoStop}%
\bibitem [{\citenamefont {Pecker}\ \emph {et~al.}(2013)\citenamefont {Pecker},
  \citenamefont {Kuemmeth}, \citenamefont {Secchi}, \citenamefont {Rontani},
  \citenamefont {Ralph}, \citenamefont {McEuen},\ and\ \citenamefont
  {Ilani}}]{Pecker:2013p2692}%
  \BibitemOpen
  \bibfield  {author} {\bibinfo {author} {\bibfnamefont {S.}~\bibnamefont
  {Pecker}}, \bibinfo {author} {\bibfnamefont {F.}~\bibnamefont {Kuemmeth}},
  \bibinfo {author} {\bibfnamefont {A.}~\bibnamefont {Secchi}}, \bibinfo
  {author} {\bibfnamefont {M.}~\bibnamefont {Rontani}}, \bibinfo {author}
  {\bibfnamefont {D.~C.}\ \bibnamefont {Ralph}}, \bibinfo {author}
  {\bibfnamefont {P.~L.}\ \bibnamefont {McEuen}},\ and\ \bibinfo {author}
  {\bibfnamefont {S.}~\bibnamefont {Ilani}},\ }\href
  {https://doi.org/10.1038/nphys2692} {\bibfield  {journal} {\bibinfo
  {journal} {Nature Physics}\ }\textbf {\bibinfo {volume} {9}},\ \bibinfo
  {pages} {576} (\bibinfo {year} {2013})}\BibitemShut {NoStop}%
\bibitem [{\citenamefont {Ercan}\ \emph {et~al.}(2021)\citenamefont {Ercan},
  \citenamefont {Coppersmith},\ and\ \citenamefont
  {Friesen}}]{Ercan:2021p35302}%
  \BibitemOpen
  \bibfield  {author} {\bibinfo {author} {\bibfnamefont {H.~E.}\ \bibnamefont
  {Ercan}}, \bibinfo {author} {\bibfnamefont {S.~N.}\ \bibnamefont
  {Coppersmith}},\ and\ \bibinfo {author} {\bibfnamefont {M.}~\bibnamefont
  {Friesen}},\ }\href {https://doi.org/10.1103/PhysRevB.104.235302} {\bibfield
  {journal} {\bibinfo  {journal} {Phys. Rev. B}\ }\textbf {\bibinfo {volume}
  {104}},\ \bibinfo {pages} {235302} (\bibinfo {year} {2021})}\BibitemShut
  {NoStop}%
\bibitem [{\citenamefont {Penthorn}\ \emph {et~al.}(2020)\citenamefont
  {Penthorn}, \citenamefont {Schoenfield}, \citenamefont {Edge},\ and\
  \citenamefont {Jiang}}]{Penthorn:2020p08680}%
  \BibitemOpen
  \bibfield  {author} {\bibinfo {author} {\bibfnamefont {N.~E.}\ \bibnamefont
  {Penthorn}}, \bibinfo {author} {\bibfnamefont {J.~S.}\ \bibnamefont
  {Schoenfield}}, \bibinfo {author} {\bibfnamefont {L.~F.}\ \bibnamefont
  {Edge}},\ and\ \bibinfo {author} {\bibfnamefont {H.}~\bibnamefont {Jiang}},\
  }\href {https://doi.org/10.1103/PhysRevApplied.14.054015} {\bibfield
  {journal} {\bibinfo  {journal} {Phys. Rev. Applied}\ }\textbf {\bibinfo
  {volume} {14}},\ \bibinfo {pages} {054015} (\bibinfo {year}
  {2020})}\BibitemShut {NoStop}%
\bibitem [{\citenamefont {MacLean}\ \emph {et~al.}(2007)\citenamefont
  {MacLean}, \citenamefont {Amasha}, \citenamefont {Radu}, \citenamefont
  {Zumb{\"u}hl}, \citenamefont {Kastner}, \citenamefont {Hanson},\ and\
  \citenamefont {Gossard}}]{MacLean:2007p1499}%
  \BibitemOpen
  \bibfield  {author} {\bibinfo {author} {\bibfnamefont {K.}~\bibnamefont
  {MacLean}}, \bibinfo {author} {\bibfnamefont {S.}~\bibnamefont {Amasha}},
  \bibinfo {author} {\bibfnamefont {I.~P.}\ \bibnamefont {Radu}}, \bibinfo
  {author} {\bibfnamefont {D.~M.}\ \bibnamefont {Zumb{\"u}hl}}, \bibinfo
  {author} {\bibfnamefont {M.~A.}\ \bibnamefont {Kastner}}, \bibinfo {author}
  {\bibfnamefont {M.~P.}\ \bibnamefont {Hanson}},\ and\ \bibinfo {author}
  {\bibfnamefont {A.~C.}\ \bibnamefont {Gossard}},\ }\href
  {https://doi.org/10.1103/PhysRevLett.98.036802} {\bibfield  {journal}
  {\bibinfo  {journal} {Phys. Rev. Lett.}\ }\textbf {\bibinfo {volume} {98}},\
  \bibinfo {pages} {036802} (\bibinfo {year} {2007})}\BibitemShut {NoStop}%
\bibitem [{\citenamefont {DiCarlo}\ \emph {et~al.}(2004)\citenamefont
  {DiCarlo}, \citenamefont {Lynch}, \citenamefont {Johnson}, \citenamefont
  {Childress}, \citenamefont {Crockett},\ and\ \citenamefont
  {Marcus}}]{Dicarlo:2004p1440}%
  \BibitemOpen
  \bibfield  {author} {\bibinfo {author} {\bibfnamefont {L.}~\bibnamefont
  {DiCarlo}}, \bibinfo {author} {\bibfnamefont {H.~J.}\ \bibnamefont {Lynch}},
  \bibinfo {author} {\bibfnamefont {A.~C.}\ \bibnamefont {Johnson}}, \bibinfo
  {author} {\bibfnamefont {L.~I.}\ \bibnamefont {Childress}}, \bibinfo {author}
  {\bibfnamefont {K.}~\bibnamefont {Crockett}},\ and\ \bibinfo {author}
  {\bibfnamefont {C.~M.}\ \bibnamefont {Marcus}},\ }\href
  {https://doi.org/10.1103/PhysRevLett.92.226801} {\bibfield  {journal}
  {\bibinfo  {journal} {Phys. Rev. Lett.}\ }\textbf {\bibinfo {volume} {92}},\
  \bibinfo {pages} {226801} (\bibinfo {year} {2004})}\BibitemShut {NoStop}%
\bibitem [{\citenamefont {Hanson}\ \emph {et~al.}(2007)\citenamefont {Hanson},
  \citenamefont {Kouwenhoven}, \citenamefont {Petta}, \citenamefont {Tarucha},\
  and\ \citenamefont {Vandersypen}}]{Hanson:2007p1217}%
  \BibitemOpen
  \bibfield  {author} {\bibinfo {author} {\bibfnamefont {R.}~\bibnamefont
  {Hanson}}, \bibinfo {author} {\bibfnamefont {L.~P.}\ \bibnamefont
  {Kouwenhoven}}, \bibinfo {author} {\bibfnamefont {J.~R.}\ \bibnamefont
  {Petta}}, \bibinfo {author} {\bibfnamefont {S.}~\bibnamefont {Tarucha}},\
  and\ \bibinfo {author} {\bibfnamefont {L.~M.~K.}\ \bibnamefont
  {Vandersypen}},\ }\href {https://doi.org/10.1103/RevModPhys.79.1217}
  {\bibfield  {journal} {\bibinfo  {journal} {Rev. Mod. Phys.}\ }\textbf
  {\bibinfo {volume} {79}},\ \bibinfo {pages} {1217} (\bibinfo {year}
  {2007})}\BibitemShut {NoStop}%
\bibitem [{\citenamefont {{v}an~{d}er Wiel}\ \emph {et~al.}(2003)\citenamefont
  {{v}an~{d}er Wiel}, \citenamefont {De~{F}ranceschi}, \citenamefont
  {Elzerman}, \citenamefont {Fujisawa}, \citenamefont {Tarucha},\ and\
  \citenamefont {Kouwenhoven}}]{vanderWiel:2003p1}%
  \BibitemOpen
  \bibfield  {author} {\bibinfo {author} {\bibfnamefont {W.~G.}\ \bibnamefont
  {{v}an~{d}er Wiel}}, \bibinfo {author} {\bibfnamefont {S.}~\bibnamefont
  {De~{F}ranceschi}}, \bibinfo {author} {\bibfnamefont {J.~M.}\ \bibnamefont
  {Elzerman}}, \bibinfo {author} {\bibfnamefont {T.}~\bibnamefont {Fujisawa}},
  \bibinfo {author} {\bibfnamefont {S.}~\bibnamefont {Tarucha}},\ and\ \bibinfo
  {author} {\bibfnamefont {L.~P.}\ \bibnamefont {Kouwenhoven}},\ }\href
  {https://doi.org/10.1103/RevModPhys.75.1} {\bibfield  {journal} {\bibinfo
  {journal} {Rev. Mod. Phys.}\ }\textbf {\bibinfo {volume} {75}},\ \bibinfo
  {pages} {1} (\bibinfo {year} {2003})}\BibitemShut {NoStop}%
\bibitem [{\citenamefont {Elzerman}\ \emph {et~al.}(2003)\citenamefont
  {Elzerman}, \citenamefont {Hanson}, \citenamefont {Greidanus}, \citenamefont
  {Willems~van Beveren}, \citenamefont {De~Franceschi}, \citenamefont
  {Vandersypen}, \citenamefont {Tarucha},\ and\ \citenamefont
  {Kouwenhoven}}]{Elzerman:2003p728}%
  \BibitemOpen
  \bibfield  {author} {\bibinfo {author} {\bibfnamefont {J.~M.}\ \bibnamefont
  {Elzerman}}, \bibinfo {author} {\bibfnamefont {R.}~\bibnamefont {Hanson}},
  \bibinfo {author} {\bibfnamefont {J.~S.}\ \bibnamefont {Greidanus}}, \bibinfo
  {author} {\bibfnamefont {L.~H.}\ \bibnamefont {Willems~van Beveren}},
  \bibinfo {author} {\bibfnamefont {S.}~\bibnamefont {De~Franceschi}}, \bibinfo
  {author} {\bibfnamefont {L.~M.~K.}\ \bibnamefont {Vandersypen}}, \bibinfo
  {author} {\bibfnamefont {S.}~\bibnamefont {Tarucha}},\ and\ \bibinfo {author}
  {\bibfnamefont {L.~P.}\ \bibnamefont {Kouwenhoven}},\ }\href
  {https://doi.org/10.1103/PhysRevB.67.161308} {\bibfield  {journal} {\bibinfo
  {journal} {Phys. Rev. B}\ }\textbf {\bibinfo {volume} {67}},\ \bibinfo
  {pages} {161308} (\bibinfo {year} {2003})}\BibitemShut {NoStop}%
\bibitem [{\citenamefont {Kittel}\ \emph {et~al.}(1980)\citenamefont {Kittel},
  \citenamefont {Charles~Kittel}, \citenamefont {Charles}, \citenamefont
  {Kroemer},\ and\ \citenamefont {Herbert}}]{Kittel:1980p10882}%
  \BibitemOpen
  \bibfield  {author} {\bibinfo {author} {\bibfnamefont {C.}~\bibnamefont
  {Kittel}}, \bibinfo {author} {\bibfnamefont {H.}~\bibnamefont
  {Charles~Kittel}}, \bibinfo {author} {\bibfnamefont {K.}~\bibnamefont
  {Charles}}, \bibinfo {author} {\bibfnamefont {H.}~\bibnamefont {Kroemer}},\
  and\ \bibinfo {author} {\bibfnamefont {K.}~\bibnamefont {Herbert}},\
  }\href@noop {} {\emph {\bibinfo {title} {Thermal Physics}}}\ (\bibinfo
  {publisher} {W. H. Freeman},\ \bibinfo {year} {1980})\BibitemShut {NoStop}%
\bibitem [{\citenamefont {Stopa}(1996)}]{Stopa:1996p13767}%
  \BibitemOpen
  \bibfield  {author} {\bibinfo {author} {\bibfnamefont {M.}~\bibnamefont
  {Stopa}},\ }\href@noop {} {\bibfield  {journal} {\bibinfo  {journal} {Phys.
  Rev. B}\ }\textbf {\bibinfo {volume} {54}},\ \bibinfo {pages} {13767}
  (\bibinfo {year} {1996})}\BibitemShut {NoStop}%
\bibitem [{\citenamefont {Wuetz}\ \emph {et~al.}(2020)\citenamefont {Wuetz},
  \citenamefont {Losert}, \citenamefont {Tosato}, \citenamefont {Lodari},
  \citenamefont {Bavdaz}, \citenamefont {Stehouwer}, \citenamefont {Amin},
  \citenamefont {Clarke}, \citenamefont {Coppersmith}, \citenamefont {Sammak},
  \citenamefont {Veldhorst}, \citenamefont {Friesen},\ and\ \citenamefont
  {Scappucci}}]{Wuetz:2020p02305}%
  \BibitemOpen
  \bibfield  {author} {\bibinfo {author} {\bibfnamefont {B.~P.}\ \bibnamefont
  {Wuetz}}, \bibinfo {author} {\bibfnamefont {M.~P.}\ \bibnamefont {Losert}},
  \bibinfo {author} {\bibfnamefont {A.}~\bibnamefont {Tosato}}, \bibinfo
  {author} {\bibfnamefont {M.}~\bibnamefont {Lodari}}, \bibinfo {author}
  {\bibfnamefont {P.~L.}\ \bibnamefont {Bavdaz}}, \bibinfo {author}
  {\bibfnamefont {L.}~\bibnamefont {Stehouwer}}, \bibinfo {author}
  {\bibfnamefont {P.}~\bibnamefont {Amin}}, \bibinfo {author} {\bibfnamefont
  {J.~S.}\ \bibnamefont {Clarke}}, \bibinfo {author} {\bibfnamefont {S.~N.}\
  \bibnamefont {Coppersmith}}, \bibinfo {author} {\bibfnamefont
  {A.}~\bibnamefont {Sammak}}, \bibinfo {author} {\bibfnamefont
  {M.}~\bibnamefont {Veldhorst}}, \bibinfo {author} {\bibfnamefont
  {M.}~\bibnamefont {Friesen}},\ and\ \bibinfo {author} {\bibfnamefont
  {G.}~\bibnamefont {Scappucci}},\ }\href
  {https://doi.org/10.1103/PhysRevLett.125.186801} {\bibfield  {journal}
  {\bibinfo  {journal} {Phys. Rev. Lett.}\ }\textbf {\bibinfo {volume} {125}},\
  \bibinfo {pages} {186801} (\bibinfo {year} {2020})}\BibitemShut {NoStop}%
\end{thebibliography}%


\begin{thebibliography}{10}
\expandafter\ifx\csname url\endcsname\relax
  \def\url#1{\texttt{#1}}\fi
\expandafter\ifx\csname urlprefix\endcsname\relax\def\urlprefix{URL }\fi
\providecommand{\bibinfo}[2]{#2}
\providecommand{\eprint}[2][]{\url{#2}}

\bibitem{Simmons:2010p245312}
\bibinfo{author}{Simmons, C.~B.} \emph{et~al.}
\newblock \bibinfo{title}{Pauli spin blockade and lifetime-enhanced transport
  in a {S}i/{S}i{G}e double quantum dot}.
\newblock \emph{\bibinfo{journal}{Phys. Rev. B}} \textbf{\bibinfo{volume}{82}},
  \bibinfo{pages}{245312} (\bibinfo{year}{2010}).

\bibitem{Penthorn:2020p08680}
\bibinfo{author}{Penthorn, N.~E.}, \bibinfo{author}{Schoenfield, J.~S.},
  \bibinfo{author}{Edge, L.~F.} \& \bibinfo{author}{Jiang, H.}
\newblock \bibinfo{title}{Direct measurement of electron intervalley relaxation
  in a $\mathrm{Si}/\mathrm{Si}$-$\mathrm{Ge}$ quantum dot}.
\newblock \emph{\bibinfo{journal}{Phys. Rev. Applied}}
  \textbf{\bibinfo{volume}{14}}, \bibinfo{pages}{054015}
  (\bibinfo{year}{2020}).

\bibitem{Elzerman:2004p731}
\bibinfo{author}{Elzerman, J.~M.}, \bibinfo{author}{Hanson, R.},
  \bibinfo{author}{Willems~van Beveren, L.~H.}, \bibinfo{author}{Vandersypen,
  L. M.~K.} \& \bibinfo{author}{Kouwenhoven, L.~P.}
\newblock \bibinfo{title}{Excited-state spectroscopy on a nearly closed quantum
  dot via charge detection}.
\newblock \emph{\bibinfo{journal}{Appl. Phys. Lett.}}
  \textbf{\bibinfo{volume}{84}}, \bibinfo{pages}{4617--4619}
  (\bibinfo{year}{2004}).

\bibitem{MacLean:2007p1499}
\bibinfo{author}{MacLean, K.} \emph{et~al.}
\newblock \bibinfo{title}{Energy-dependent tunneling in a quantum dot}.
\newblock \emph{\bibinfo{journal}{Phys. Rev. Lett.}}
  \textbf{\bibinfo{volume}{98}}, \bibinfo{pages}{036802}
  (\bibinfo{year}{2007}).

\bibitem{Dicarlo:2004p1440}
\bibinfo{author}{DiCarlo, L.} \emph{et~al.}
\newblock \bibinfo{title}{Differential charge sensing and charge delocalization
  in a tunable double quantum dot}.
\newblock \emph{\bibinfo{journal}{Phys. Rev. Lett.}}
  \textbf{\bibinfo{volume}{92}}, \bibinfo{pages}{226801}
  (\bibinfo{year}{2004}).

\bibitem{Hanson:2007p1217}
\bibinfo{author}{Hanson, R.}, \bibinfo{author}{Kouwenhoven, L.~P.},
  \bibinfo{author}{Petta, J.~R.}, \bibinfo{author}{Tarucha, S.} \&
  \bibinfo{author}{Vandersypen, L. M.~K.}
\newblock \bibinfo{title}{Spins in few-electron quantum dots}.
\newblock \emph{\bibinfo{journal}{Rev. Mod. Phys.}}
  \textbf{\bibinfo{volume}{79}}, \bibinfo{pages}{1217--1265}
  (\bibinfo{year}{2007}).

\bibitem{vanderWiel:2003p1}
\bibinfo{author}{{v}an~{d}er Wiel, W.~G.} \emph{et~al.}
\newblock \bibinfo{title}{Electron transport through double quantum dots}.
\newblock \emph{\bibinfo{journal}{Rev. Mod. Phys.}}
  \textbf{\bibinfo{volume}{75}}, \bibinfo{pages}{1--22} (\bibinfo{year}{2003}).

\bibitem{Elzerman:2003p728}
\bibinfo{author}{Elzerman, J.~M.} \emph{et~al.}
\newblock \bibinfo{title}{Few-electron quantum dot circuit with integrated
  charge read out}.
\newblock \emph{\bibinfo{journal}{Phys. Rev. B}} \textbf{\bibinfo{volume}{67}},
  \bibinfo{pages}{161308} (\bibinfo{year}{2003}).

\bibitem{Kittel:1980p10882}
\bibinfo{author}{Kittel, C.}, \bibinfo{author}{Charles~Kittel, H.},
  \bibinfo{author}{Charles, K.}, \bibinfo{author}{Kroemer, H.} \&
  \bibinfo{author}{Herbert, K.}
\newblock \emph{\bibinfo{title}{Thermal Physics}} (\bibinfo{publisher}{W. H.
  Freeman}, \bibinfo{year}{1980}).

\bibitem{Stopa:1996p13767}
\bibinfo{author}{Stopa, M.}
\newblock \bibinfo{title}{Quantum dot self-consistent electronic structure and
  the coulomb blockade}.
\newblock \emph{\bibinfo{journal}{Phys. Rev. B}} \textbf{\bibinfo{volume}{54}},
  \bibinfo{pages}{13767--13783} (\bibinfo{year}{1996}).

\bibitem{Wuetz:2020p02305}
\bibinfo{author}{Wuetz, B.~P.} \emph{et~al.}
\newblock \bibinfo{title}{Effect of quantum hall edge strips on valley
  splitting in silicon quantum wells}.
\newblock \emph{\bibinfo{journal}{Phys. Rev. Lett.}}
  \textbf{\bibinfo{volume}{125}}, \bibinfo{pages}{186801}
  (\bibinfo{year}{2020}).

\bibitem{Ercan:2021p35302}
\bibinfo{author}{Ercan, H.~E.}, \bibinfo{author}{Coppersmith, S.~N.} \&
  \bibinfo{author}{Friesen, M.}
\newblock \bibinfo{title}{Strong electron-electron interactions in
  {S}i/{S}i{G}e quantum dots}.
\newblock \emph{\bibinfo{journal}{Phys. Rev. B}}
  \textbf{\bibinfo{volume}{104}}, \bibinfo{pages}{235302}
  (\bibinfo{year}{2021}).

\end{thebibliography}
%\bibliography{/Users/jpdodson/Repositories/UWbibfiles/bib-files/main.bib}

\end{document}

\def\simlt{\mathrel{\lower .3ex \rlap{$\sim$}\raise .5ex \hbox{$<$}}}

\title{\textbf{\fontfamily{phv}\selectfont 
How valley-orbit states in silicon quantum dots probe quantum well interfaces}}
\author{J. P. Dodson}
\affiliation{Department of Physics, University of Wisconsin-Madison, Madison, WI 53706, USA}
\author{H. Ekmel Ercan}
\affiliation{Department of Physics, University of Wisconsin-Madison, Madison, WI 53706, USA}
\author{J. Corrigan}
\affiliation{Department of Physics, University of Wisconsin-Madison, Madison, WI 53706, USA}
\author{Merritt P. Losert}
\affiliation{Department of Physics, University of Wisconsin-Madison, Madison, WI 53706, USA}
\author{Nathan Holman}
\affiliation{Department of Physics, University of Wisconsin-Madison, Madison, WI 53706, USA}
\author{Thomas McJunkin}
\affiliation{Department of Physics, University of Wisconsin-Madison, Madison, WI 53706, USA}
\author{L. F. Edge}
\affiliation{HRL Laboratories, LLC, 3011 Malibu Canyon Road, Malibu, CA 90265, USA}
\author{Mark Friesen}
\affiliation{Department of Physics, University of Wisconsin-Madison, Madison, WI 53706, USA}
\author{S. N. Coppersmith}
\affiliation{Department of Physics, University of Wisconsin-Madison, Madison, WI 53706, USA}
\affiliation{University of New South Wales, Sydney, Australia}|
\author{M. A. Eriksson}
\affiliation{Department of Physics, University of Wisconsin-Madison, Madison, WI 53706, USA}

\maketitle
\startsupplement

\section{Pulsed-gate spectroscopy model}

\begin{figure}[t]
\vspace{.1in} 
\includegraphics[width=0.48\textwidth]{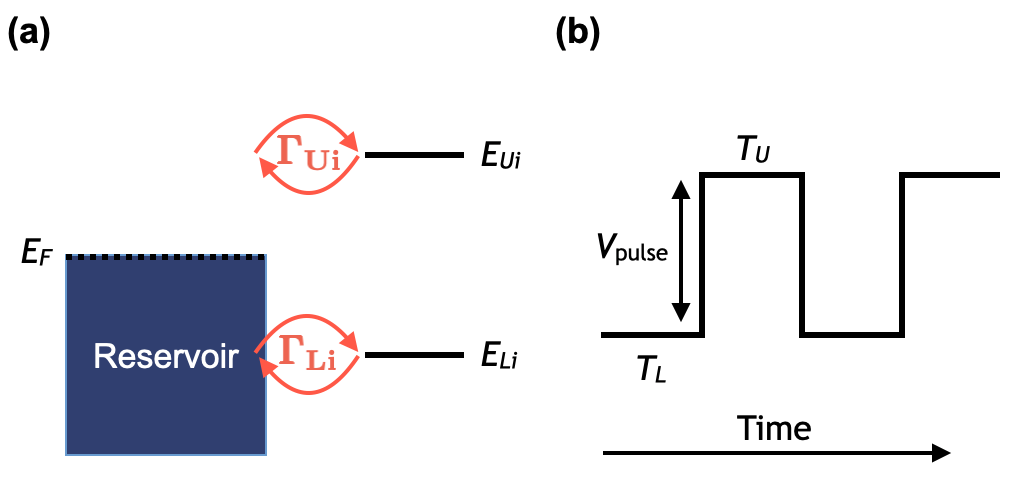}
\caption{
\label{fig:figS2}
Parameters definitions for pulse spectroscopy.
(a) $E_F$ is the Fermi energy of the reservoir. 
$E_{Ui}$ is the chemical potential of the dot eigenstate state $i$ at the top edge of the pulse.
Here, $i$ can refer to the ground state ($g$), the first excited state ($x_1$), etc.
$E_{bi}$ is the chemical potential of the dot eigenstate $i$ at the bottom edge of the pulse.
$\Gamma_{Ui}$ and $\Gamma_{Li}$ are the corresponding tunnel couplings.
(b) The duty cycle for the pulse applied to gate P2. $T_L = T_U$ for valley-orbit measurements. The pulse frequency $f_{\text{pulse}} = 1/(T_L + T_U)$ is on the order of the tunnel rate into/out of the dot during the bottom part of the pulse cycle. The tunnel rate out of the dot during the top part of the pulse is much faster than the pulse frequency such that the dot is completely empty at the beginning of the bottom part of the pulse. 
}
\end{figure}

Consider a dot tunnel coupled to a reservoir, illustrated in Fig.~\ref{fig:figS2}(a).
The dot is pulsed in a two-step duty cycle between voltages $V_L$ and $V_U$, as shown in Fig.~\ref{fig:figS2}(b).
The corresponding chemical potentials of the ground states of the dot, $E_{Lg}$ and $E_{Ug}$, are related to $V_L$ and $V_U$ via a lever arm: $E_{\nu g} = \alpha V_{\nu} + E_0$, where $\nu=U,L$ and $E_0$ is the ground state (g) of the dot in the absence of a pulse.
If the dot also has an excited state (x), then $\Delta E\equiv E_{Ux}-E_{Ug}=E_{Lx}-E_{Lg}$, and $E_{\nu x}=E_{\nu g}+\Delta E$.
Tunneling can occur if there is a state in the reservoir that electrons can tunnel out of or into, as determined by the Fermi function:
\begin{equation}
f_{\nu i}=\left[ e^{(E_{\nu i}-E_F)/k_BT}+1\right]^{-1} ,
\end{equation}
where $i=g,x$.

The tunnel rates, $\Gamma_{\nu i}$, depend on the chemical potential $E_{\nu i}$, and can differ for the different dot eigenstates ($\nu=g,x$).
As discussed in Ref.~\cite{Simmons:2010p245312}, $\Gamma_{\nu i}$ is a complicated exponential function.
However, since there is not enough structure in the data to fully characterize the features of $\Gamma_{\nu i}$, we can simply linearize this function about the Fermi level, obtaining
\begin{gather}
\Gamma_{\nu g} \simeq \Gamma_{0,\nu g} e^{(E_{\nu g}-E_F)/E_{0,\nu g}} , \\
\Gamma_{\nu x} \simeq \Gamma_{0,\nu x} e^{(E_{\nu g}+\Delta E-E_F)/E_{0,\nu x}} .
\end{gather}
While there also exists a decay rate from the excited state to the ground state, $\Gamma_{xg}$, here we do not consider its effect since recent measurements \cite{Penthorn:2020p08680} indicate the decay rate is much slower than our pulse frequency $f_{\text{pulse}} \simeq $ 1 MHz.

For a single ground state in the quantum dot, the rate equation for its level filling $n_g$ is derived using detailed balance:

\begin{equation}
\label{eq:detailed_balance}
\dot{n}_{\nu g} = \Gamma_{\nu g}(f_{\nu g} - n_{\nu g})
\end{equation}

Before solving Eqn.~\ref{eq:detailed_balance}, we note that the steady state filling of the ground state, $n_g$, can be computed by setting $\dot{n}_g = 0$. The steady state filling is then simply the Fermi function, $n_g = f_g$. However, this is only valid in the limit that $\Gamma_{\nu g} \gg f_{\text{pulse}}$. The measurements shown in the main text tune the tunnel rates such that, $\Gamma \simeq f_{\text{pulse}}$, giving high sensitivity for pulsed gate spectroscopy measurements \cite{Elzerman:2004p731}. Thus, we must find the general solution to Eqn.~\ref{eq:detailed_balance} for $t > t_0$:

\begin{equation}
n_{\nu g}(t) = f_{\nu g} - f_{\nu g}\left(1 - \dfrac{n_{\nu g}(t_0)}{f_{\nu g}}\right)e^{-\Gamma_{\nu g}(t-t_0)}
\end{equation}

\noindent where $n_{\nu g}(t_0)$ is the occupation of the dot at $t = t_0$. Since $\alpha V_{\text{pulse}}$ is much larger than the singlet-triplet splitting ($E_{\text{ST}}$) and valley splitting ($E_{\text{val}}$),  $\Gamma_{Ug} \gg f_{\text{pulse}}$ in the fit regime \cite{MacLean:2007p1499}. Thus, when fitting our data, it is assumed that the dot is completely empty at the beginning of the bottom part of the duty cycle. Then $n_{Lg}(t_0) = 0$ and Eqn.~\ref{eq:detailed_balance} simplifies to

\begin{equation}
\label{eq:nbg_general_sol}
n_{Lg}(t) = f_{Lg}\left(1 - e^{-\Gamma_{Lg}(t-t_0)}\right)
\end{equation}

\noindent The average electron occupation of the ground state of the dot during the entire pulse cycle is found by integrating from $t=t_0$ to $t=t_0 + T_L$ and dividing by $1/(T_U + T_L)$, where it has been assumed the average electron occupation of the dot during the top part of the duty cycle is 0 and $T_U = T_L$. Integrating Eqn.~\ref{eq:nbg_general_sol}, we obtain

\begin{equation}
\label{eq:navg_pulsed_gate}
\braket{n_{g}} = \dfrac{1}{2}f_{Lg} + \dfrac{1}{2T_L}\dfrac{f_{Lg}}{\Gamma_{Lg}}\left(e^{-\Gamma_{Lg}T_L} - 1\right)
\end{equation}

\noindent where the first term on the right hand side of Eqn.~\ref{eq:navg_pulsed_gate} represents the steady state occupation of the dot during the bottom part of the pulse, and the second term represents the perturbation due to energy dependent tunneling \cite{Simmons:2010p245312}. Finally, assuming that $\braket{n_L} = \braket{n_{Lg}} + \braket{n_{Lx}}$, the average electron occupation for the ground and excited states in the dot during the bottom part of the pulse is

\begin{equation}
\label{eq:navg_pulsed_gate_final_general}
\braket{n} = \dfrac{1}{2}\sum \limits_{i = g, x} \left[f_{Li} + \dfrac{1}{T_L}\dfrac{f_{Li}}{\Gamma_{Li}}\left(e^{-\Gamma_{Li}T_L} - 1\right)\right]
\end{equation}

\noindent This form can be simplified further into a convenient fitting form by expanding the exponential in Eqn.~\ref{eq:navg_pulsed_gate_final_general}. This is valid for our measurements since $\Gamma \sim f_{\text{pulse}}$ and $T_L = 1/(2f_{\text{pulse}})$, making $\Gamma_{Li} T_L < 1$. Expanding to second order, we obtain the final fitting form for determining valley-orbit state splittings using pulsed-gate spectroscopy

\begin{equation}
\label{eq:navg_pulsed_gate_final}
\braket{n} = \sum \limits_{i = g,x} \Gamma_i \dfrac{e^{(E_i-E_{F})/E_{0i}}}{e^{(E_i-E_F)/k_B T_e} + 1}
\end{equation}
%\braket{n_{b}} = \dfrac{T_b}{4}\sum \limits_{i = g, x} f_{bi}\Gamma_{bi}

\noindent where $E_i = \alpha V_i$ is the position of each peak in energy, and $E_{0i}$ and $\Gamma_i$ are fitting parameters for each peak.

\section{Magnetospectroscopy model}
\label{sec:magspec}

Magnetospectroscopy was used in addition to pulsed-gate spectroscopy to supplement measurements of $E_{\text{ST}}$, providing better precision at low valley-orbit splittings and quantitative verification of pulsed-gate valley-orbit splitting measurements. In this section, a model is developed that can be used to extract $E_{\text{ST}}$ from magnetospectroscopy data sets, where the plunger gate voltage is swept on one axis and the magnetic field is swept on the other. This 2-dimensional data set is then converted into a 1-dimensional data set by peak-fitting along the voltage axis, where the differential conductance is fit using the derivative of of Eqn.~2 in Ref.~\cite{Dicarlo:2004p1440} with respect to gate voltage. The peak-fitted data set is then curve fit using Eqn.~\ref{eq:fitVg}, which we now derive.

Consider a quantum dot that is tuned electrostatically such that the one- and two-electron charge states are nearly degenerate. The system is found to be in one of six spin states that is dependent on the number of electrons in the quantum dot. For one electron, it is found to be in the $\ket{\uparrow}$ or $\ket{\downarrow}$ spin-state. For two-electrons, it is found to be in one of the singlet ($\ket{S_0}$) or triplet ($\ket{T_-}, \ket{T_0}, \ket{T_+}$) spin-states. Including charge and spin degrees of freedom and excluding charging energy, the energy of each state is given by \cite{Hanson:2007p1217, vanderWiel:2003p1}:
\begin{align}
E_{\downarrow} &= - \dfrac{1}{2}g\mu_B B - \alpha V_{\text{P2}} \label{downspineqn}\\
E_{\uparrow} &= \dfrac{1}{2}g\mu_B B - \alpha V_{\text{P2}}  \\
E_{S_0} &= - 2\alpha V_{\text{P2}} \\ 
E_{T_-} &= - g\mu_B B - 2\alpha V_{\text{P2}} + E_{\text{ST}} \\
E_{T_0} &= - 2\alpha V_{\text{P2}} + E_{\text{ST}} \\
E_{T_+} &= g\mu_B B - 2\alpha V_{\text{P2}} + E_{\text{ST}} \label{Tpluseqn}
\end{align}

\noindent where $g=1.99$ is the electron g-factor in silicon, $\mu_B$ is the Bohr magneton, $\alpha$ is the gate lever arm, and $V_{\text{P2}}$ is the gate voltage. 

Using standard lock-in detection techniques, detailed in Ref.~\cite{Elzerman:2003p728}, the derivative of the average electron occupation of the dot with respect to gate voltage is measured, shown in Fig.~\ref{fig:fig2}(e) of the main text. A derivation for the average electron occupation of a system in thermal and diffusive contact with a large reservoir is detailed in Ref.~\cite{Kittel:1980p10882}. The total electron occupation $\braket{n}$ is determined by treating the system as a grand canonical ensemble:
\begin{align}
\braket{n} &=\dfrac{1}{\mathcal{Z}}\sum\limits_i N_i e^{(N\mu - E_i)/k_BT_e} \label{eq:grandpartition} \\
\mathcal{Z} &= \sum\limits_i e^{(N_i\mu - E_i)/k_BT_e}
\end{align}
where $\mathcal{Z}$ is the grand partition function, $\mu$ is the chemical potential of the reservoir, $T_e$ is the electron temperature and $N_i$ is the number of electrons on the dot for each given state (two electrons for the $\ket{S_0}$ and $\ket{T}$ states, and one electron for $\ket{\uparrow}$ and $\ket{\downarrow}$ states).

A 4-state model (inclusion of states $\ket{\downarrow}, \ket{\uparrow}, \ket{S_0}, \ket{T_-}$) may be sufficient in the limit of low $T_e$ and high $E_{\text{ST}}$; however, we find that in the experimentally relevant range ($T_e \simeq 100$ mk, $E_{\text{ST}} = $ 25--60 $\mu$eV), the 6-state model is requisite since significant occupation of higher lying states at $B = 0$ occurs. We do not attempt to quantitatively match the raw voltage $V_{\text{P2}}$, instead plotting $\delta V_{\text{P2}}$. This simplifies the final result (Eqn.~\ref{eq:grandpartition}), allowing  $\mu = 0$ as this quantity only provides an offset in $V_{\text{P2}}$. Additionally, this justifies excluding the charging energy from the states at the beginning of the derivation. The average electron occupation is then

\begin{widetext}
\begin{align}
\braket{n} = 1 + \dfrac{e^{\alpha \beta_e V_{\text{P2}}}\left(1 + e^{\kappa B}\left(1 + e^{E_{\text{ST}}\beta_e}+e^{\kappa B}\right)\right)}{e^{\alpha \beta_e V_{\text{P2}}} + e^{(\alpha V_{\text{P2}} + E_{\text{ST}})\beta_e + \kappa B} + e^{E_{\text{ST}}\beta_e + \frac{1}{2}\kappa B} + e^{\alpha \beta_e V_{\text{P2}} + \kappa B} + e^{E_{\text{ST}} \beta_e + \frac{3}{2} \kappa B} + e^{\alpha \beta_e V_{\text{P2}} + 2\kappa B}} 
 \label{eq:navg_magspec}
 \end{align}
\end{widetext}
%\dfrac{d\braket{N}}{dV_g} = \frac{\alpha  \beta  \left(e^{\beta  B g \mu }+1\right) \left(e^{\beta  B g \mu } \left(e^{\beta  B g \mu }+e^{\beta  \text{ST}}+1\right)+1\right) e^{\frac{1}{2} \beta 
   %B g \mu +\beta  \text{ST}+\alpha  \beta  V}}{\left(e^{\frac{1}{2} \beta  B g \mu +\beta  \text{ST}}+e^{\frac{3}{2} \beta  B g \mu +\beta  \text{ST}}+e^{\beta  (B g
   %\mu +\text{ST}+\alpha  V)}+e^{\beta  B g \mu +\alpha  \beta  V}+e^{2 \beta  B g \mu +\alpha  \beta  V}+e^{\alpha  \beta  V}\right)^2}
\noindent where we have introduced $\beta_e = 1/k_B T_e$ and $\kappa=g \mu_B \beta_e$. The derivative of Eqn.~\ref{eq:navg_magspec} with respect to gate voltage is plotted in Fig.~\ref{fig:fig2}(f) of the main text.

When fitting for $E_{\text{ST}}$, the peak value position along the gate voltage axis ($V_{\text{P2}}$) is extracted as described above (ie: Fig.~\ref{fig:fig2}(g) of the main text). The curve fit is obtained from \ref{eq:navg_magspec} by solving the equation $d^2\braket{n}/dV_{\text{P2}}^2 = 0$. The resulting curve fit as a function of magnetic field is

\begin{align}
 V_{P2}(B) = \dfrac{1}{\alpha\beta} \ln \left(\dfrac{e^{\frac{1}{2} \kappa B + \beta_e E_{\text{ST}}}\left(e^{\kappa B}+1\right)}{e^{\kappa B}+e^{2 \kappa B}+e^{\kappa B + \beta_e E_{\text{ST}}}+1}\right), \label{eq:fitVg}
\end{align}

\section{Comparison between pulsed-gate spectroscopy and magnetospectroscopy}
\label{magpulse}

Fig.~\ref{fig:fig4} of the main text reports values of $E_{\text{ST}}$ and $E_{\text{val}}$ measured using both magnetospectroscopy and pulsed-gate spectroscopy, so it is important to confirm quantitative agreement between the two measurement techniques. Pulsed-gate spectroscopy is the more powerful of the two techniques for determining valley-orbit state splittings since single and multi-electron valley-orbit states can be detected. However, in some cases, excited-state signals are too weak due to tunnel rate constraints on the system, making magnetospectroscopy the superior method. This is especially true in regimes where valley-orbit energies approach the electron temperature $T_e$. For measurements here, it became difficult to distinguish between the ground and excited state signals below $\sim$ 35 $\mu$eV due to thermal broadening. 

\begin{figure}
\begin{centering}
\includegraphics[width=0.48\textwidth]{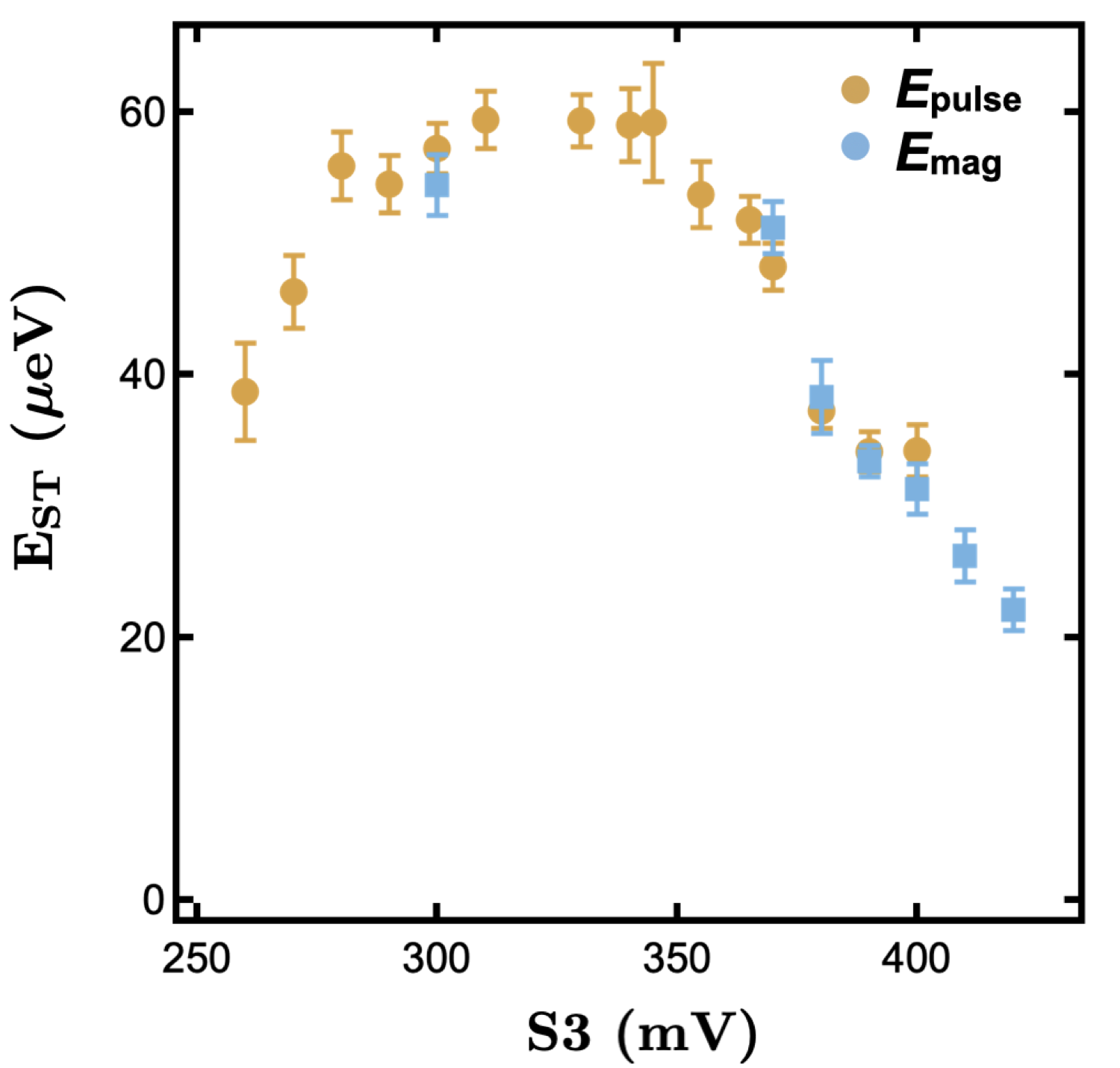}
\caption{\label{fig:fig5_3} Comparison of singlet-triplet splitting $E_{\text{ST}}$ measured at different electrostatic configurations. The screening gate voltage $V_{\text{S3}}$ changes the electrostatic confinement potential of the dot. $E_{\text{ST}}$ is measured as a function of the $V_{\text{S3}}$ using both pulsed-gate spectroscopy (red) and magnetospectroscopy (blue). Good agreement is found between the two methods, where magnetospectroscopy is used for smaller values of $E_{\text{ST}}$ since pulsed-gate spectroscopy has difficulty resolving two distinct peaks below 35 $\mu$eV at an electron temperature of 100 mK.}
\end{centering}
\end{figure} 

Fig.~\ref{fig:fig5_3} shows the comparison of $E_{\text{ST}}$ measured by magnetospectroscopy (blue squares) and pulsed-gate spectroscopy (orange circles). Magnetospectroscopy data was not taken at all electrostatic configurations due to the length of the measurement. Since magnetospectroscopy was the primary method used for low $E_{\text{ST}}$, we emphasized measurements in that regime. Agreement within the error bars (given as the standard deviation) is found between the two methods.

\section{COMSOL electrostatic simulations}
\label{sec:comsol}

\renewcommand{\arraystretch}{1.3}
\begin{table}[t]
\caption{\label{table:table2} COMSOL gate voltages used for reaching a symmetric, single-electron dot beneath P2. The reservoir beneath gate R2 is extended beneath plunger gate P1 such that P2 is tunnel coupled to the R2 reservoir. P3 is depleted such that tunneling into the R1 reservoir is cut off. } 
\centering\begin{tabular*}{\columnwidth}{@{\extracolsep{\stretch{1}}}*{7}{c}@{}}
\hline\hline 
& & \\[-2ex]
 & Gate Name & COMSOL Voltage (mV)\\ [1ex] \hline	
 	& P1 & 450  \\
    & P2 & 575 \\
 	& P3 & 580 \\
 	& B1 & 450 \\
 	& B2 & 450 \\
 	& B3 & 450 \\
 	& B4 & 580 \\
 	& S1 & 445 \\
 	& S2 & 405\\
 	& S3 & 425\\
 	& R1 & 650\\
 	& R2 & 650\\
 	& R3 & 650\\
 	& R4 & 650\\
 	& T1 & 470\\
 	& T2 & 490\\
 	& M1 & 590\\
	\hline\hline
\end{tabular*}
\end{table}

The COMSOL Multiphysics simulations are full device simulations where the dimensions for each gate electrode are modeled out to 500 nm away from the center of the device. Using a realistic quasiclassical (Thomas-Fermi) electrostatic simulation \cite{Stopa:1996p13767, Wuetz:2020p02305}, the electron density is calculated as a function of gate voltages. The simulation is used to gain intuition and provide estimates for quantum dot shape and position as the electrostatic configuration is varied across the experimental range.

The Thomas-Fermi simulations are made to be as consistent as possible with experimental device tune-up by using the following procedure: First, all voltages in the simulation are ramped to the exact same voltages as the device. Next, the presence of offset charge and the workfunction difference between materials is incorporated into the simulation by adding an offset voltage until the P2 quantum dot is in the single-electron regime (as it is in experiment). This is analogous to adding a global offset in the threshold voltage of the device. At this point, all differential gate voltages from the experimental configuration have been preserved. One additional offset is made to reach the starting configuration---a single electron in symmetric quantum dot beneath the P2 gate: All screening gates are increased by 55 mV, which gives a symmetric quantum dot beneath P2 in the COMSOL simulation. By only choosing a single offset from the experimental voltages, this preserves the differential voltages between screening gates, which are the primary confinement gates along the y-axis. For reference, the gate voltages used in the COMSOL simulation to achieve the single-electron regime in Fig.~\ref{fig:fig1} of the main text are shown in Table \ref{table:table2}.

The final step in simulating the experimental range of electrostatic tunings is adjusting $V_{\text{S3}}$ and the gates neighboring the P2 quantum dot to closely approximate the gate voltage changes made in the experiment. The gate lever arms ($\alpha$) are determined for each gate in the COMSOL simulation, and are found to be in close agreement (but not identical) to the experiment. Table \ref{table:table1} shows the experimentally measured $\alpha$ for each gate neighboring the P2 quantum dot. To keep the P2 quantum dot electron occupation close to 1.0 throughout the different electrostatic regimes, COMSOL offset voltages were determined using the ratio of experimental to COMSOL gate lever arm values as $\Delta V_{\text{sim}} = \alpha_{\text{exp}}/\alpha_{\text{sim}} \Delta V_{\text{exp}}$. By applying offsets in this way, the differential gate voltage used in the experiment is preserved without significantly changing the electron occupation of the simulated dot.

The COMSOL simulated orbital splittings and change in position of the quantum dot qualitatively match the experimental results, giving insight into the physical behavior of the device across the electrostatic configurations. Both the simulation and measurements indicate $\hbar \omega_x < \hbar \omega_y$ at low S3 gate voltages, and $\hbar \omega_x > \hbar \omega_y$ at large S3 gate voltages. Between the S3 voltages of 330--370 mV, the dot is found to be roughly isotropic in the simulation. 

As noted in the main text, when simulating experimental voltages of $V_{\text{S3}} > 400$ mV, physical results were not produced because of the onset of electron accumulation directly beneath the S3 screening. Between $400-420$ mV, the P2 quantum dot becomes unphysically elongated along the y-axis, and at $>420$ mV, nonzero accumulation beneath the S3 gate is observed. The highest S3 gate voltage value where an orbital splitting was able to be measured in experiment was at 430 mV, and at 440 mV, accumulation beneath the S3 gate occurred. This gives an estimated difference of $20$ mV between the experiment and theoretical simulations for the upper bound of the S3 gate voltage. The discrepancy is most likely due to a small difference is $\alpha_{\text{S3}}$ between the simulation and experiment since the $\alpha$ of each gate varies across the electrostatic tunings, as shown in Table \ref{table:table1}.

\renewcommand{\arraystretch}{1.3}
\begin{table}[t]
\caption{\label{table:table1} Measurements of $\alpha$ for gates S3, S2, P2, B2 and B3 at varying electrostatic configurations (denoted by the S3 gate voltage value). In addition to the changing S3 gate voltage, a compensating voltage to neighboring plunger and barrier gates was applied to maintain the electron occupation of the P2 dot. The change in $\alpha$ values clearly shows the dot moving away from S2 towards S3 as the S3 gate voltage changes from 260 mV to 420 mV. This exact behavior observed in the COMSOL simulations.} 
\centering\begin{tabular*}{\columnwidth}{@{\extracolsep{\stretch{1}}}*{7}{r}@{}}
\hline\hline 
 &  &  &  &  &  \\[-2ex]
S3 (mV) & $\quad$ $\alpha_{S3}$ & $\quad$ $\alpha_{S2}$ & $\quad$ $\alpha_{P2}$ & $\quad$ $\alpha_{B2}$ & $\quad$ $\alpha_{B3}$ \\ [1ex] \hline	
	260 & 0.231 & 0.263 & 0.209 & 0.063 & 0.098 \\
	330 & 0.226 & 0.213 & 0.207 & 0.064 & 0.089 \\
	370 & 0.230 & 0.179 & 0.198 & 0.061 & 0.079 \\
	420 & 0.277 & 0.134 & 0.189 & 0.064 & 0.075 \\
	\hline\hline
\end{tabular*}
\end{table}

\section{TB and FCI methods}
\label{sec:FCI}
%\begin{figure}[t]
%\includegraphics[width=0.50\textwidth]{figS8.png}
%\caption{\label{fig:figS8} Atomic step profile used for FCI calculations and corresponding dot position as a function of electrostatic tuning. (a) An atomic step profile with a constant step density of 35.16 nm/step corresponding to a miscut of 0.2$^\circ$ of the substrate is used to simulate valley-orbit coupling induced by the atomic details of the heterostructure quantum well interface. A search is performed for each experimental data point in Fig.~\ref{fig:fig3}(b) of the main text, where the optimal position of the quantum dot relative to the interface is extracted from FCI calculations. Valley and singlet-triplet splittings are calculated using FCI at each red line shown in (a), and the point that matches experiment best is extracted and plotted in (b) as black circles. (b) Dot position plotted as distance from interface step as a function of electrostatic tuning. The data set is fit linearly to extract a total change in position of 23.7 nm.} 
%\end{figure} 
Full configuration interaction (FCI) simulations aided by tight-binding (TB) calculations of the electronic wave functions are used to accurately model $E_{\text{val}}$ and $E_{\text{ST}}$, as shown in Fig.~\ref{fig:fig3} of the main text. The TB model calculations results in wave functions with fast oscillations in the z-direction which breaks the valley degeneracy. Thus, effects on the valley splitting due to atomistic disorder, and the disorder-induced valley-orbit coupling are captured in the TB model. FCI methods allow for accurate calculation of two-electron energies through diagonalization of the interacting two-electron Hamiltonian in the basis of 3160 Slater determinants generated by using the TB solutions. Therefore, the combination of these two procedures capture the interplay among the valleys, interface disorder and e-e interactions.

The FCI calculations use a number of experimentally driven input parameters. This includes the measured $E_{\text{orb}}$ for each point, a vertical electric field of 0.7 MV m$^{-1}$, a 9 nm thick quantum well and atomic steps at the quantum well interface separated by 35 nm. The confinement potential of the quantum dot is based derived from $E_{\text{orb}}$, which is used in the form of a harmonic potential for determining the electron wave function in the x-y plane of the quantum dot. Additionally, this influences the strength of e-e interactions. The procedure for simulating valley and singlet-triplet energies is detailed more thoroughly in Ref.~\cite{Ercan:2021p35302}.

For the calculation of quantum dot position with respect to the position of atomic steps at the quantum well interface, the following procedure is used: first, the singlet-triplet and valley splittings are simulated as the quantum dot position changes. The dot position is varied in steps of 1.76 nm such that a matrix of 20 values are simulated between two atomic steps separated by 35 nm. Next, the simulated values are compared to experimental values, and the point which simultaneously has the smallest total error between simulation and experiment for both singlet-triplet and valley splittings is chosen. Finally, the dot is only allowed to move in one direction, otherwise the results would be unphysical. This condition is enforced by noting that the simulations produce a reflectional symmetry halfway between the steps. Two low-error simulations are found on both sides of this symmetry point. Thus, once the quantum dot has moved past the symmetry point, it cannot move backwards across it.

\bibliography{main.bib}
%\bibliography{/Users/jpdodson/Repositories/UWbibfiles/bib-files/main.bib}